\documentclass{aa}  

\usepackage{algorithm,algpseudocode} 
\usepackage{graphicx}
\usepackage{natbib} 
\usepackage{mathcomp}
\usepackage{txfonts}
\begin{document}

   \title{Variable Star Classification Using Multi-View Metric Learning}
   \titlerunning{Variable Star Classification Using Multi-View Metric Learning}

   \author{K. B. Johnston \inst{1}\inst{5}
            S.M. Caballero-Nieves, \inst{1}
            V. Petit, \inst{2}
            A.M. Peter, \inst{3}
            \and R. Haber, \inst{4}}

\authorrunning{K. B. Johnston et al.}
   \institute{Aerospace, Physics and Space Sciences Dept., Florida Institute of
Technology, 150 W. University Blvd., Melbourne, FL, US\\
              \email{kyjohnst2000@my.fit.edu}
         \and
             Physics and Astronomy Dept., University of Delaware, Newark, DE, USA,
             \and
             Computer Engineering and Sciences Dept., Florida Institute of Technology,
150 W. University Blvd., Melbourne, FL, US
\and
Mathematical Sciences Dept., Florida Institute of Technology, 150
W. University Blvd., Melbourne, FL, US
\and
Defense Group Melbourne, Perspecta Inc., 4849 N. Wickham Rd., Melbourne, FL, USA
             }

 
  \abstract
   {Comprehensive observations of variable stars can include time domain
photometry in a multitude of filters, spectroscopy, estimates of color
(e.g. U-B), etc. When the objective is to classify variable stars, traditional
machine learning techniques distill these various representations (or views) into 
a single feature vector and attempt to discriminate among desired categories.}
   {In this work,
we propose an alternative approach that inherently leverages multiple views of the same
variable star. }
   {Our multi-view metric learning framework enables robust characterization
of star categories by directly learning to discriminate in a multi-faceted feature space, thus,
eliminating the need to combine feature representations prior to fitting the 
machine learning model.  We also demonstrate how to extend standard multi-view
learning, which employs multiple vectorized views, to the matrix-variate case which 
allows very novel variable star signature representations.}
   {The performance of our proposed
methods is evaluated on the UCR Starlight and LINEAR datasets.  Both the vector and matrix-variate versions
of our multi-view learning framework perform favorably --- demonstrating the ability to
discriminate variable star categories..}
   {}


   \maketitle
%

\section{Introduction}
The classification of variable stars relies on the proper selection of tell-tale light-curve
signatures representing a specific variability type (referred to as the features of 
interest in machine learning terminology) and an automated detection/separation framework that
can differentiate different variable stars (referred to as the classifier in machine learning
terminology).

Classically, the astroinformatics-community standard features include: the quantification
of statistics associated with the time domain photometric data (e.g., mean, standard deviation, kurtosis), Fourier
decomposition of the data (e.g., ratio of frequency, peak frequency), and color information in both the optical
and infrared domain \citep{nun2015fats,miller2015machine}. 

Likewise, standard classification techniques in the astro\-informatics-community span a few areas: 
(i) the classifier is designed such that the user selects features and the classifier is trained on variables with a known type \citep["expert selected features, for correlation discovery", ][]{Debosscher2009,Sesar2011,Richards2012,graham2013machine,armstrong2015k2,mahabal2017deep,hinners2018machine},

(ii) the classifier is designed such that the computer selects the optimal features and the classifier is trained on variables with a known type \citep["computer selected features, for correlation discovery"]{mcwhirter2017classification,naul2018recurrent},

(iii) the classifier (clustering algorithm) is designed such that that user selects features and variables with an unknown type are provided \citep["expert selected features, for class discovery", ][]{valenzuela2017unsupervised,modak2018unsupervised}.

These efforts have been hampered by multiple factors. First, the underlying foundational data to be used in classification is biased either resulting
from the original composition of the survey from which the training data is selected
\citep{Angeloni2014} or the choice of building a training set containing only a subset of the 
top five to ten most frequent class-types \citet{kim2016package, Pashchenko2018, naul2018recurrent}. 
In our research, no efforts were found in the literature that address all variables 
identified by \citet{kazarovets2017general}---most address some subset. 

Second, the
development of expertly selected feature sets is often keyed to the original selection of 
variable stars of interest, and their definition. As surveys become more complete
and more dense in observations, the complexity of the problem is likely to grow as 
a result of further refinement of classification definitions \citep{kazarovets2017general}. 

Third, the legacy expertly selected features \citep{richards2011machine} are often co-linear,
resulting in little to no new information or separability despite the increase in 
dimensionality and additional increase in computational power needed to manage the data \citep{d2016analysis}. 
Lastly, increases in feature dimensionality (with co-linear data) results in needlessly 
increasing the sparsity of the training data space (e.g., curse of dimensionality). This 
requires increasingly more complex classifier designs to both support the dimensionality
as well as the potential non-linear class-space separation. 

Presented here is a methodology that addresses these issues using a combination of new features and advanced classifiers designs. 

Two novel transforms, Slotted Symbolic Markov Model \citep[SSMM,][]{johnston2017variable} and  Distribution
Fields \citep[DF,][]{johnston:inpress-b}, are used to generate viable feature spaces for the classification of variable stars. SSMM requires no phasing of the time domain data but still provides a feature that is shape based, DF allows for the consideration of the whole phased waveform without additional picking and choosing of metrics from the waveform \citep[i.e., see ][]{Helfer2015}. Both of these transforms require only processed (artifact removed) time domain data, i.e. light curves.

Each of these new transforms quantify the light curve shape, using either phased and un-phased data. This methodology of quantifying all available time domain information within the transformations identified and allowing the selected classifier to optimized the features of interests, contrasts with the traditional methodology of using un-optimized, biased, sub-selected features that may---or may not---contain information that is vital to the discrimination of different types of variable stars. 

However, the features we selected are matrix-variate (i.e. $\mathbb{R}^{m\text{\texttimes}n}$), thus to accommodate the usage of either of these features (or both) we introduce two classifiers to the astroinformatics community. The first, Large Margin Multi-view Metric Learning \citep[LM\textsuperscript{3}L, ][]{hu2014large} relies on dimensionality reduction methodologies. The second, Large Margin Multi-view Metric Learning with Matrix Variates (LM\textsuperscript{3}L-MV), is a novel development and inherently generates compact feature spaces as part of the optimization process. Both classifiers allow for the usage of multiple domains in the classification process (both SSMM and DF simultaneously). 

Multi-view learning is a methodology that can provide a major benefit to the astronomical community. Astronomy often deals with multiple transformations (e.g., Fourier Domain, Wavelet Domain, statistical...etc) and multiple domains of data types (visual, radio frequency, high energy, particle, etc.). The ability to handle, and just as importantly co-train an optimization algorithm on, multiple domain data will be necessary as the multitude of data grows. Furthermore, metric learning decisions have an implicit traceability: the ability to follow from the classifier's decision, to the weights associated with each individual feature used as part of the classification, to the nearest-neighbors used in making the decision provide a clear idea of why the classifier made the decision. This direct comparison of newly observed with prior observations, and the justification via historical comparison, make this method ideal for astronomical---and indeed scientific---applications. 

This paper additionally outlines a system design that allows for the tools provided here to be translated to many different scenarios, using many different input values, providing interested scientists flexibility of use.  
In this paper will be organized as follows: (1) summarize current stellar variable classification efforts, features currently in use, and machine learning methodologies exercised (2) review the features used (statistics, color, DF and SSMM) (3) review the classification
methodologies used (metric learning, LM\textsuperscript{3}L, and LM\textsuperscript{3}L-MV)
(4) demonstrate our optimization of feature extraction algorithm for
our datasets, leveraging ``simple'' classification methods (k-NN) and cross-validation processes (5) demonstrate
our optimization of classification parameters for LM\textsuperscript{3}L and LM\textsuperscript{3}L-MV
via cross-validation and (6) report on the performance of the feature/classifier
pairing. Our proposal is an implementation of both the feature extraction
and classifier for the purposes of multi-class identification, that
can handle raw observed data. The project software is provided publicly at the associated GitHub
repository \footnote{https://github.com/kjohnston82/VariableStarAnalysis}

\section{Theory and Design}

We present an initial set of time domain feature extraction methods;
the design demonstrated is modular in nature, allowing for a user to append or
substitute feature spaces that an expert has found to be of utility
in the identification of variable stars. Although our initial goal is variable star identification, given a separate set of features this method could be
applied to other astroinformatics problems (i.e., image classification
for galaxies, spectral identification for stars or comets, etc.).
While we demonstrate the classifier has a multi-class classification
design, which is common in the astroinformatics references we have
provided, the design here can easily be transformed into a one-vs-all
design \citep{johnston2017generation} for the purposes of generating
a detector or classifier designed specifically to a user's needs \citep{johnston:inpress-b}.

\subsection{Signal Conditioning}

Required are features that can respond to the various signal
structures that are unique to the classes of interest, i.e. phased light shape, frequency distribution, phase distribution, etc.). Our implementation starts with raw data (such as astronomical light curves) as primary input, which are then mapped into a specific feature space. To support these transformations,
a set of signal conditioning methods are implemented for the two new feature space presented below. These techniques
are based on the methods presented in \citet{johnston2017variable}
and are fairly common in the industry. The data that is leveraged---
with respect to classification of the waveform---is on the order of
hundreds of observations over multiple cycles. While the data is not
cleaned as part of the upfront process, the features that are
implemented are robust enough to not be affected by intermittent
noise. The raw waveform is left relatively unaffected, however smoothing
does occur on the phased waveform to generate a new feature vector,
i.e. a phased smoothed waveform. 

\begin{figure*}

\centering
\begin{minipage}{.5\textwidth}
  \centering
  \includegraphics[scale=0.2]{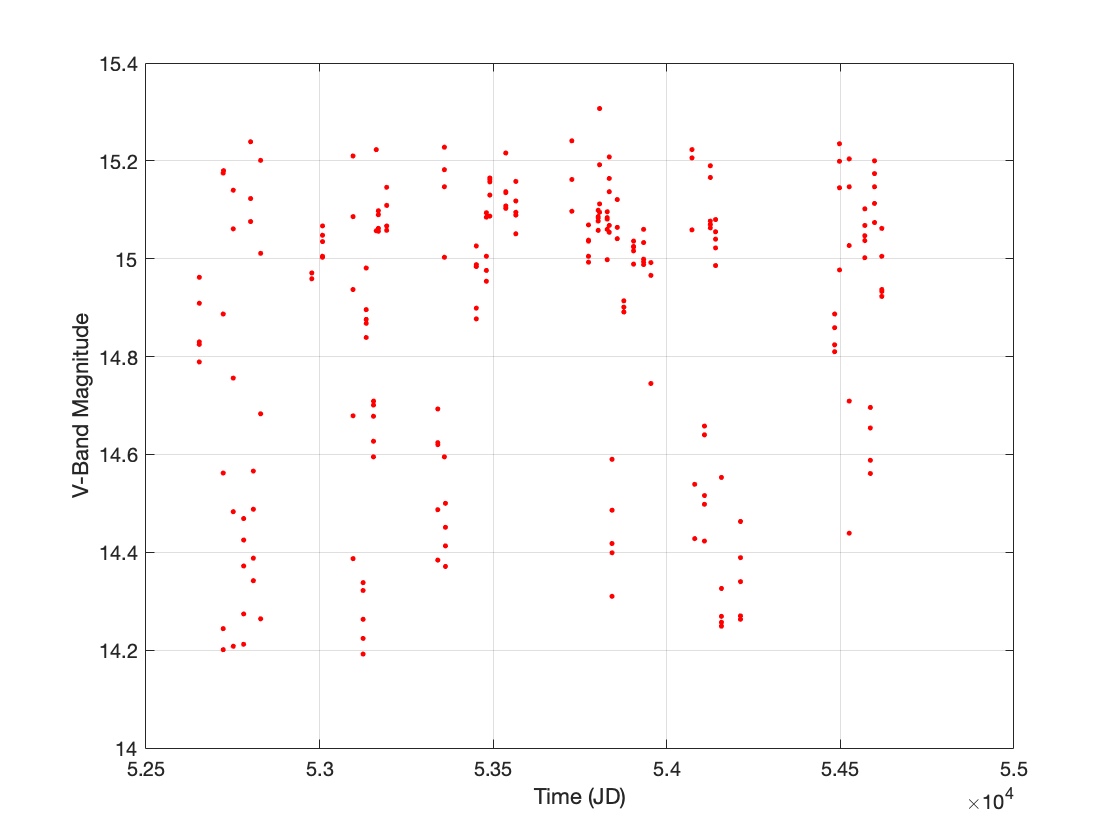}
\end{minipage}%
\begin{minipage}{.5\textwidth}
  \centering
  \includegraphics[scale=0.2]{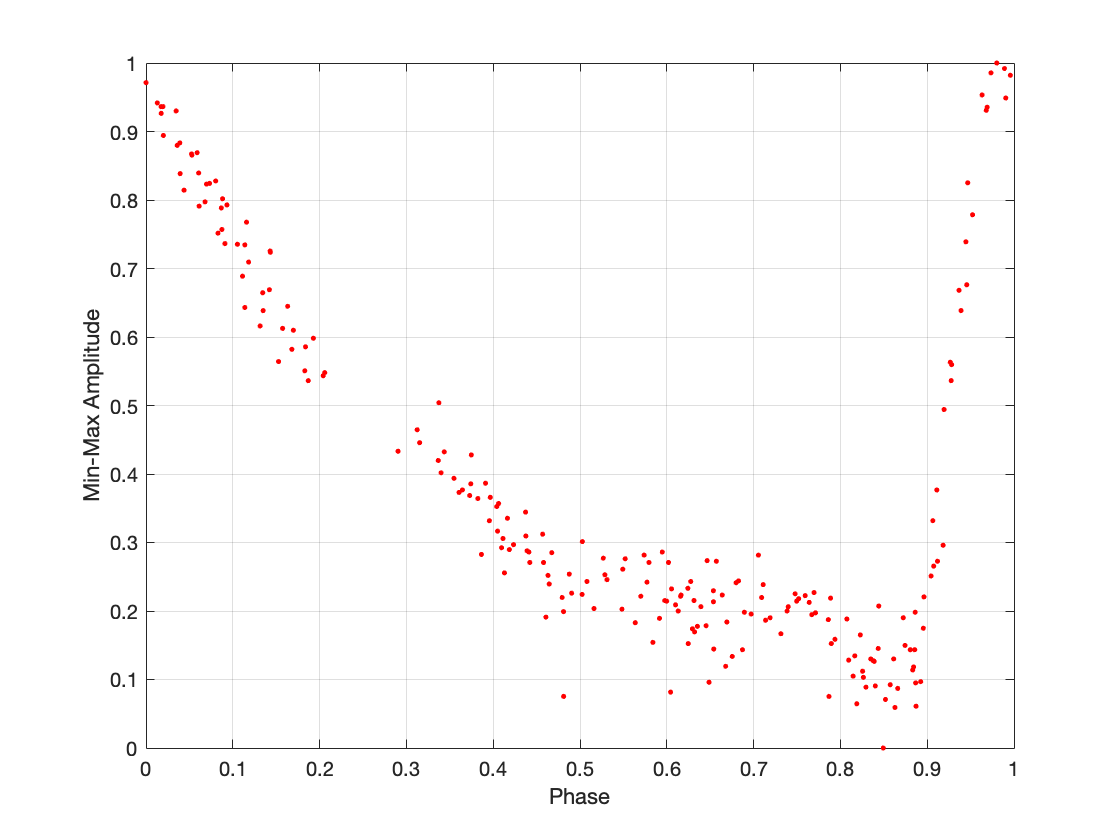}
\end{minipage}
\caption{Left: Raw Time Domain Data, Right: Corresponding Transformed Phased Time Domain Data, Example Given for RR Lyr Type Variable}
\label{fig:Phased-time-domain}
\end{figure*}

The phased waveform is generated via folding the raw waveform about a period found to best represent
the cyclical process \citep{graham2013comparison}. The SUPER-SMOOTHER
algorithm \citep{friedman1984variable} is used to smooth the phased data into a functional representation. Additionally in some cases, the originating survey/mission will perform some of these signal conditioning processes as part of their analysis pipeline (e.g., Kepler). This includes outlier removal, period finding, and long term trend removal. Most major surveys include a processing pipeline, our modular analysis methods provide a degree of flexibility that allow the implementer to take advantage of these pre-applied processes. Specifically of use, while our feature extraction SSMM does not require a phased waveform, the DF feature does, thus period finding methods are of importance.

Most of the period finding algorithms are methods of spectral
transformation with an associated peak/max/min finding algorithm and
include such methods as: discrete Fourier transform, wavelets decomposition, least squares approximations, string length, auto-correlation, conditional
entropy and auto-regressive methods. \citet{graham2013comparison}
review these transformation methods (with respect to period finding),
and find that the optimal period finding algorithm is different for
different types of variable stars. The Lomb--Scargle method was selected as the main method for generating a primary period for this implementation. For more information, our implementation
of the Lomb--Scargle algorithm is provided as part of the Variable
Star package\footnote{fit.astro.vsa.analysis.feature.LombNormalizedPeriodogram}.

\subsection{Feature Extraction}

For our investigation we have selected feature spaces that quantify
the functional shape of repeated signal---cyclostationary signal---but are dynamic enough to handle impulsive type signals (e.g., supernova) as well. 
This particular implementation design makes the most intuitive sense,
visual inspection of the light curve is how experts identify these sources. Prior research on time domain data identification
has varied between generating machine learned features \citep{Bos2002,Broersen2009,Blomme2011,Bolos2014,gagniuc2017markov},
implementing generic features \citep[e.g. Fourier domain features; ][]{Debosscher2009,Richards2012,Palaversa2013,Masci2014},
and looking at shape or functional based features \citep[e.g. DF, SSMM; ][]{Park2013,haber2015discriminative}.

We implement two novel time domain feature space transforms: SSMM
and DF. It is not suggested that these features are going to be the
best in all cases, nor are they the only choice as is apparent from
\citet{fulcher2013highly}. Any feature space, so long as it provides
separability, would be usable here. One need only think of how to
transform the observable (time domain, color, spectra, etc.) into
something that is a consistent signature for stars in given class-type (i.e., variable star type).

SSMM itself is an effective feature for discriminating
variable star types as shown by \citet{johnston2017variable}. Similarly, DF has been shown to be a valuable feature for discriminating time domain signatures, see \citet{Helfer2015} and \citet{johnston:inpress-b}.

\subsubsection{Slotted Symbolic Markov Models (SSMM)}

Slotted Symbolic Markov Models (SSMM) is useful
in the differentiation between variable star types \citep{johnston2017variable}.
The time domain slotting described in \citet{rehfeld2011comparison}
is used to regularize the sampling of the photometric observations.
The resulting regularized sampled waveform is transformed into a state
space \citep{lin2007experiencing,bass2016supervised}; thus the result
of the conditioning is the stochastic process $\left\{ y_{n},\:n=1,2,...\right\} $.
The stochastic process is then used to populate the empty matrix $\mathbf{P}$
\citep{Ge2000}---the elements of $\mathbf{P}$ are populated as the
transition state probabilities (equation \ref{eq:1}).

\begin{equation}
\mathbf{P}\{y_{n+1}=j|\:y_{n}=i,\:y_{n-1}=i_{n-1},...,\:y_{1}=i_{1},\:y_{0}=i_{0}\}=P_{ij}\label{eq:1}
\end{equation}

The populated matrix $\mathbf{P}$ is the SSMM feature; and is often
described as a first order Markov Matrix. 

\subsubsection{Distribution Field (DF)}

A distribution field (DF) is an array of probability distributions,
where probability at each element is defined as equation \ref{eq:2} \citep{Helfer2015, sevilla2012distribution}.

\begin{equation}
DF_{ij}=\frac{\sum_k^N\left[y_{j}<f\left(x_{i}\leq\phi_k\leq x_{i+1}\right)_N<y_{j-1}\right]}{\sum_k^N\left[y_{j}<f\left(\phi_k\right)_N<y_{j-1}\right]},\label{eq:2}
\end{equation}
where N is the number of samples in the phased data, and  $\left[\:\right]$ is the Iverson bracket \citep{iverson1962programming},
given as
\begin{equation}
[P]=\left\{ \begin{array}{cc}
1 & P={\rm true}\\
0 & {\rm otherwise,}
\end{array}\right.\label{eq:iverson}
\end{equation}
Similarly, $f\left(\phi_k\right)$ is the smoothed phased data, and $y_{j}$ and $x_{i}$ are the corresponding normalized amplitude
and phase bins. The bins are defined as $x_i = {0, 1/n_x, 2/n_x, \dots, 1}$ and
$y_i = {0, 1/n_y, 2/n_y, \dots, 1}$; $n_x$ is the number of time
bins, and $n_y$ is the number of amplitude
bins. The result is a right stochastic matrix, i.e., the rows sum to 1. 
Bin number, $n_x$ and $n_y$, is optimized by cross-validation as
part of the classification training process.

\begin{figure}
\centering{}\includegraphics[scale=0.2]{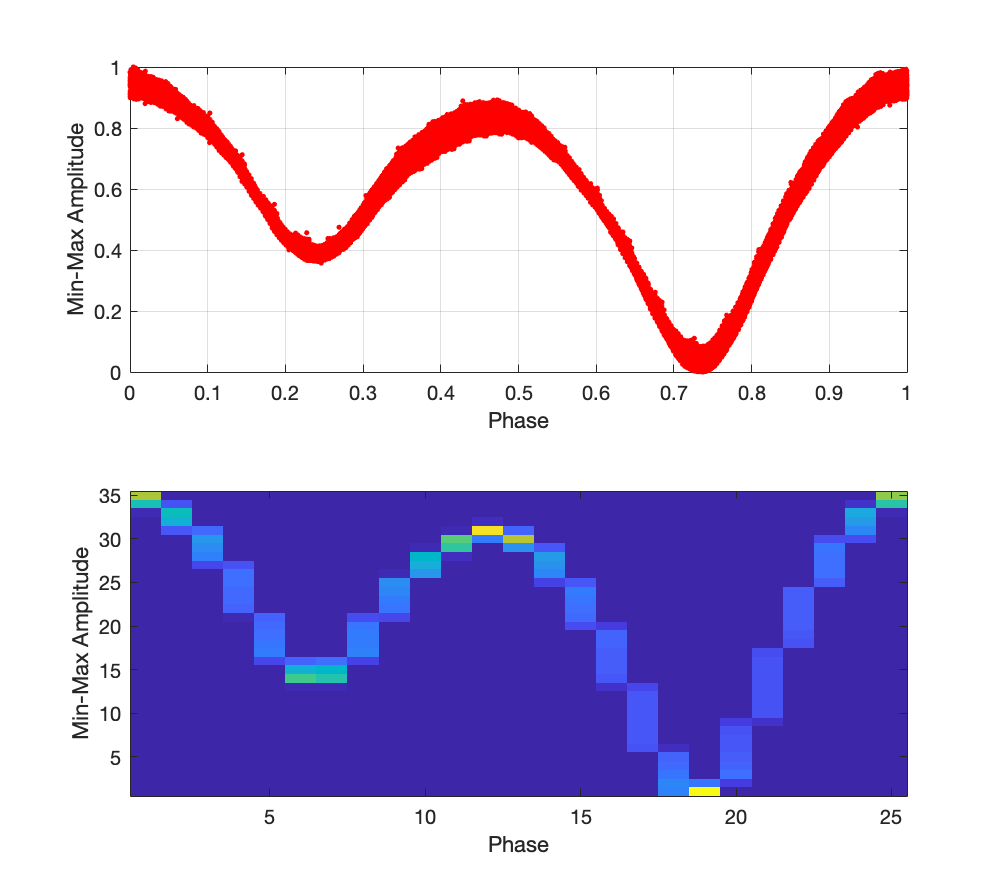}
\caption{Folded Time Domain Data Transformed into the DF Feature space, Example Given for Eclipsing Binary Type Variable\label{fig:DF-Example}}
\end{figure}

\subsection{Classification and Metric Learning}

The classification methodology known as metric learning has its roots
in the understanding of how and why observations are considered similar.
The very idea of similarity is based around the numerical measurement
of distance, and the computation of a distance is generated via application
of a distance function. \citet{bellet2015metric} define the metric distance as
equation \ref{eq:3}

\begin{equation}
d(x,x')=\sqrt{\left(x-x'\right)^{T}\mathbf{M}\left(x-x'\right)}\label{eq:3};
\end{equation}

where $X\subseteq\mathbb{R}^{d}$ and the metric is required to be
$\mathbf{M}\in\mathbb{S}_{+}^{d}$. $\mathbb{S}_{+}^{d}$ is
the cone of symmetric positive semi-definite (PSD) $d\times d$ real
valued matrices. Metric learning seeks to optimize this distance,
via manipulation of the metric $\mathbf{M}$, based on available side
data. How the optimization occurs, what is focused on and what is
considered important, i.e. the construction of the objective function,
is the underlying difference between the various metric learning algorithms. 

The side information is defined as the set of labeled data
$\left\{ x_{i},y_{i}\right\} _{i=1}^{n}$. Furthermore the triplet
is defined as $\left(x_{i},x_{j},x_{k}\right)$ where $x_{i}$ and $x_{j}$
have the same label but $x_{i}$ and $x_{k}$ do not. It is expected
then, based on the definition of similarity and distance, that $d(x_{i},x_{j})<d(x_{i},x_{k})$,
i.e., that the distances between similar labels is smaller than the
distances between dissimilar ones. Methods such as Large Margin Nearest Neighbor \citep[LMNN, ][]{weinberger2006distance}
use this inequality to defined an objective function that optimizes the metric to bring similar things closer together, while pushing dissimilar things further apart. 

Given the metric learning optimization process, the result is a tailored
distance metric and associated distance function (equation \ref{eq:3}).
This distance function is then used in a standard k-NN classification
algorithm. The k-NN algorithm estimates a classification label based
on the closest samples provided in training \citep{altman1992introduction}.
If ${x_{n}}$ is a set of training data $n$ big, then we find the
distance between a new pattern $x_{i}$ and each pattern in the training
set. The new pattern is classified depending on the majority of class
labels in the closest $k$ data points. 

Prior studies have initially addressed the potential of using metric
learning as a means for classification of variable stars \citep{johnston:inpress-b}.
Metric learning has a number of benefits that are advantageous to
the astronomer:
\begin{itemize}
\item Metric learning uses nearest neighbors (k-NN) classification to generate the decision
space \citep{Hastie2009,Duda2012}, k-NN provides instant clarity
into the reasoning behind the classifiers decision (based on similarity,
``$x_{i}$ is closer to $x_{j}$ than $x_{k}$'' ).
\item Metric learning leverages side information (the supervised labels
of the training data) to improve the metric, i.e. a transformation
of the distance between points that favors a specific goal: similar
closer together, different further apart, simplicity of the metric,
feature dimensionality reduction, etc.. This side data is based on
observed prior analyzed data, thus decisions have a grounding in expert
identification as opposed to black-box machine learning \citep{bellet2015metric}.
Dimensionality reduction in particular can be helpful for handling
feature spaces that are naturally sparse.
\item k-NN can be supported by other algorithm structures such as data partitioning
methods to allow for a rapid response time in assigning labels to
new observations, despite relying upon a high number of training data
\citep{faloutsos1994fast}.
\item The development of an anomaly detection functionality \citep{Chandola2009},
which has been shown to be necessary to generate meaningful classifications
\citep[see:][]{johnston2017variable,johnston2017generation}, is easily
constructed from the k-NN metric learning framework.
\end{itemize}

\subsubsection{Multi-View Learning}

We address the following classification problem: given a set of
expertly labeled side data containing $C$ different classes (e.g., variable star types), where
measurements can be made on the classes in question to extract a set
of feature spaces for each observation, how do we define a distance metric
that optimizes the misclassification rate? Specifically, how can this
be done within the context of variable star classification based on
the observation of photometric time-domain data? We have identified a number of features that provide utility in discriminating
between various types of stellar variables. We review how to combine this information
together and generate a decision space; or rather,
how to define the distance $d_{ij}=(x_{i}-x_{j})'\mathbf{M}(x_{i}-x_{j})$,
when $x_{i}$ contains two matrices (SSMM or DF in our case). Specifically
we attempt to construct a distance metric based on multiple attributes
of different dimensions (e.g. $\mathbb{R}^{m\text{\texttimes}n}$
and $\mathbb{R}^{m\times1}$ ). 

To respond to this challenge we investigate
the utility of multi-view learning. For our purposes here we specify
each individual measurement as the feature, and the individual transformations or representations of the underlying measurement as the feature space. Views, are the generic independent collections of these features or
feature space. Thus, if provided the color of a variable star in \textit{ugriz},
the individual measurements of $u-g$ or $r-i$ shall be referred
to here as the features but the collective set of colors is the
feature space. Our methodology here allows us to defined sets of collections
of these feature and/or feature spaces as independent views, for example:
all of \textit{ugriz} measurements, the vectorized DF measurement, the concatenation
of time-domain statistics and colors, the reduced (selected) sampling
of Fourier spectra, could all be individual views. The expert defined these views {\it a priori}.

\cite{xu2013survey} and \cite{kan2016multi} review multi-view learning and outline some
basic definitions. Multi-view learning treats the individual views
separately, but also provides some functionality for joint learning
where the importance of each view is dependent on the others. As an
alternative to multi-view learning, the multiple views could be transformed
into a single view, usually via concatenation. The cost--benefit analysis
of concatenated single-view vs. multi-view learning are discussed in \citet{xu2013survey}
and are beyond the scope of this paper. 

Classifier fusion \citep{kittler1998combining,Tax2001,Tax2003}
could be considered as an alternative to multi-view learning, with each
view independently learned, and resulting in an independent classification
algorithm. The result of the set of these classifiers are combined
together (mixing of posterior probability) to result in a singular
estimate of classification/label; this is similar to the operation
of a Random Forest classifier, i.e. results from multiple individual
trees combined together to form a joint estimate. We differentiate between the single-view
learning with concatenation, multi-view learning, and classifier fusion
designs based on when the join of the views is considered in the optimization process: before, during, or after.

Multi-view learning can be divided into three topics:
1) co-training, 2) multiple-kernel learning, and 3) subspace learning.
Each method attempts to consider all views during the training process.
Multiple-kernel learning algorithms attempt to exploit kernels that
naturally correspond to different views and combine kernels either
linearly or non-linearly to improve learning performance \citep{gonen2011multiple, kan2016multi}.

Sub-space learning uses canonical correlation analysis (CCA), or a
similar method, to generate an optimal latent representation of two
views which can be trained on directly. The CCA method can be iterated multiple times based on the number of views, this process will frequently
result in a dimensionality that is lower then the original space
\citep{hotelling1936relations,akaho2006kernel,zhu2012dimensionality, kan2016multi}. 

This work will focus on a method of co-training, specifically metric
co-training. Large Margin Multi-Metric Learning \citep{hu2014large,hu2017MMMM}
is an example of metric co-training; the designed objective function
minimizes the optimization of the individual view, as well as the
difference between view distances, simultaneously. The full derivation
of this algorithm is outlined in \citet{hu2014large}, and the algorithm
for optimization for LM\textsuperscript{3}L is given as their Algorithm 1. We have implemented the algorithm in Java and it is available as part of the software distribution.

Our implementation also includes additional considerations not discussed in the original
reference. These considerations were found to be necessary based on
challenges discovered when we applied the LM\textsuperscript{3}L algorithm to
our data. The challenges and our responses are discussed in Appendix \ref{sec:challenges}. 

In addition to the implementation
of LM\textsuperscript{3}L, we have developed a matrix variate version as well
(section \ref{subsec:Large-Margin-Multi-Metric}). This matrix variate classifier is novel with respect to multi-view learning methods and is one of two metric learning methods that we know of, the other being Push-Pull Metric Learning \citep{Helfer2015}.

\subsection{Large Margin Multi-View Metric Learning with Matrix Variates \label{subsec:Large-Margin-Multi-Metric}}

The literature on metric learning methods is fairly extensive (see
\citet{bellet2015metric} for a review), however all of the methods
presented so far focus on the original
definition that is based in $X\subseteq\mathbb{R}^{d\times1}$ , i.e.
vector-variate learning. While the handling of matrix-variate data
has been addressed here, the method require a transformation---$\mathrm{vec}(x)$ and then ECVA---which ignores the problem of directly operating on matrix-variate data. The literature on matrix-variate classification and operations is
fairly sparse. The idea of a metric learning supervised classification
methodology based on matrix-variate data is novel.

Most of the matrix-variate research has some roots in the work by
\citet{hotelling1936relations} and \citet{dawid1981some}. There
are some key modern references to be noted as well: \citet{ding2014dimension}
and \citet{ding2018matrix} address matrix-variate PCA and matrix variate
regression (matrix predictor and response), \citet{dutilleul1999mle}
and \citet{zhou2014gemini} address the mathematics of the matrix
normal distribution, and \citet{safayani2011matrix} address matrix-variate
CCA.

Developing a matrix-variate metric learning algorithm requires a formal
definition of distance for matrix-variate observations, i.e. where
$X\subseteq\mathbb{R}^{p\times q}$. \citet{glanz2013expectation}
define the matrix normal distribution as $X_{i}\sim MN\left(\mu,\mathbf{\Sigma}_{s},\mathbf{\Sigma}_{c}\right)$,
where $X_{i}$ and $\mu$ are $p\times q$ matrices, $\mathbf{\Sigma}_{s}$
is a $p\times p$ matrix defining the row covariance, and $\mathbf{\Sigma}_{c}$
is a $q\times q$ matrix defining the column covariance. Equivalently
the relationship between the matrix normal distribution and the vector
normal distribution is given as equation \ref{eq:23},

\begin{equation}
\mathrm{vec}\left(X_{i}\right)\sim N\left(\mathrm{vec}\left(\mu\right),\mathbf{\Sigma}_{c}\otimes\mathbf{\Sigma}_{s}\right)\label{eq:23}.
\end{equation}

The matrix-variate normal distribution is defined as equation \ref{eq:24} \citep{gupta2000matrix}

\begin{multline}
P\left(X_{i};\mu,\mathbf{\Sigma}_{s},\mathbf{\Sigma}_{c}\right)=\left(2\pi\right)^{-\frac{pq}{2}}\left|\left(\mathbf{\Sigma}_{c}\otimes\mathbf{\Sigma}_{s}\right)^{-1}\right|^{\frac{1}{2}} \\
\exp\left\{ -\frac{1}{2}\mathrm{vec}\left(X_{i}-\mu\right)\mathrm{\left(\mathbf{\Sigma}_{c}\otimes\mathbf{\Sigma}_{s}\right)^{-1}vec}\left(X_{i}-\mu\right)\right\} \label{eq:24}.
\end{multline}

This distribution holds for the features that we
are using as part of this study, at least within the individual classes. The Mahalanobis
distance between our observations is then defined for the Matrix-Variate
case as equations \ref{eq:25} to \ref{eq:27}:

\begin{equation}
d_{\mathbf{\Sigma}}(X,X')\doteq\mathrm{vec}\left(X-X'\right)\mathrm{\left(\mathbf{\Sigma}_{c}\otimes\mathbf{\Sigma}_{s}\right)^{-1}vec}\left(X-X'\right)\label{eq:25},
\end{equation}

\begin{equation}
=\mathrm{vec}\left(X-X'\right)^{T}\mathrm{vec}\left(\mathbf{\Sigma}_{s}^{-1}\left(X-X'\right)\mathbf{\Sigma}_{c}^{-1}\right)\label{eq:26},
\end{equation}

\begin{equation}
=\mathrm{tr}\left[\mathbf{\Sigma}_{c}^{-1}\left(X-X'\right)^{T}\mathbf{\Sigma}_{s}^{-1}\left(X-X'\right)\right]\label{eq:27}.
\end{equation}

This last iteration of the distance between matrices is used
in our development of a metric learning methodology. Similar to the
development of LM\textsuperscript{3}L and the outline of \citet{torresani2007large},
we develop a metric learning algorithm for matrix-variate data. First the Mahalanobis distance
for the matrix-variate multi-view case is recast as equation \ref{eq:28}

\begin{equation}
d_{\mathbf{U}_{k},\mathbf{V}_{k}}(X_{i}^{k},X_{j}^{k})=\mathrm{tr}\left[\mathbf{U}_{k}\left(X_{i}^{k}-X_{j}^{k}\right)^{T}\mathbf{V}_{k}\left(X_{i}^{k}-X_{j}^{k}\right)\right]\label{eq:28};
\end{equation}

where $\mathbf{U}_{k}$ and $\mathbf{V}_{k}$ represent the inverse
covariance of the column and row respectively. The individual view
objective function is constructed similar to the LMNN \citep{weinberger2006distance}
methodology; we define a push (equation \ref{eq:29}) and pull (equation
\ref{eq:30}) as:

\begin{multline}
push_{k}=\gamma\sum_{j\rightsquigarrow i,l}\eta_{ij}^{k}\left(1-y_{il}\right) \\
\cdot h\left[d_{\mathbf{U}_{k},\mathbf{V}_{k}}(X_{i}^{k},X_{j}^{k})-d_{\mathbf{U}_{k},\mathbf{V}_{k}}(X_{i}^{k},X_{l}^{k})+1\right]\label{eq:29},
\end{multline}

\begin{equation}
pull_{k}=\sum_{i,j}\eta_{ij}^{k}\cdot d_{\mathbf{U}_{k},\mathbf{V}_{k}}(X_{i}^{k},X_{j}^{k})\label{eq:30};
\end{equation}

where $y_{il}=1$ if and only if $y_{i}=y_{l}$ and $y_{il}=0$ otherwise;
and $\eta_{ij}^{k}=1$ if and only if $x_{i}$ and $x_{j}$ are targeted
neighbors of similar label $y_{i}=y_{j}$. For a more in-depth discussion
of target neighbor, see \citet{torresani2007large}. 

Furthermore, we include regularization terms \citep{schultz2004learning} with respect to
$\mathbf{U}_{k}$ and $\mathbf{V}_{k}$ as part of
the objective function design; these are defined as $\lambda\left\Vert \mathbf{U}_{k}\right\Vert _{F}^{2}$
and $\lambda\left\Vert \mathbf{V}_{k}\right\Vert _{F}^{2}$, respectively. The inclusion of regularization terms in our objective function help promote sparsity in the learned metrics. Favoring sparsity can be beneficial when the dimensionality of the feature spaces is high, and can help lead to a more generic and stable solution.

The sub-view objective function is then equation \ref{eq:32}:

\begin{multline}
\min_{\mathbf{U}_{k},\mathbf{V}_{k}}I_{k}=\sum_{i,j}\eta_{ij}^{k}\cdot d_{\mathbf{U}_{k},\mathbf{V}_{\mathbf{k}}}(X_{i}^{k},X_{j}^{k})+ \\
\gamma\sum_{j\rightsquigarrow i,l}\eta_{ij}^{k}\left(1-y_{il}\right)\cdot h\left[d_{\mathbf{U}_{k},\mathbf{V}_{k}}(X_{i}^{k},X_{j}^{k})-d_{\mathbf{U}_{k},\mathbf{V}_{k}}(X_{i}^{k},X_{l}^{k})+1\right]+\\ 
\lambda\left\Vert \mathbf{U}_{k}\right\Vert _{F}^{2}+
\lambda\left\Vert \mathbf{V}_{k}\right\Vert _{F}^{2}\label{eq:32};
\end{multline}

where $\lambda>0$ and controls the importance of the regularization.
From LM\textsuperscript{3}L the objective function is equation \ref{eq:33}:

\begin{multline}
\min_{\mathbf{U}_{k},\mathbf{V}_{k}}J_{k}=w_{k}I_{k}+\\ 
\mu\sum_{q=1,q\neq k}^{K}\sum_{i,j}\left(d_{\mathbf{U}_{k},\mathbf{V}_{k}}(X_{i}^{k},X_{j}^{k})-d_{\mathbf{U}_{l},\mathbf{V}_{l}}(X_{i}^{q},X_{j}^{q})\right)^{2}\label{eq:33};
\end{multline}

where $\sum_{k=1}^{K}w_{k}=1$ and the first term is the contribution
of the individual $k^{th}$ view, while the second term is designed
to minimize the distance difference between attributes. Using  these objective function definitions, we 
derive the gradient descent optimization procedure in Appendix
\ref{sec:derLM3L} following a similar procedures used by
\cite{weinberger2006distance}. The resulting methodology is 
proposed in Algorithm \ref{alg:--Algorithm}.

\begin{algorithm}[h]
\begin{centering}
\caption{LM\textsuperscript{3}L-MV Algorithm Flow \label{alg:--Algorithm}}
\par\end{centering}
\begin{algorithmic}[1] 
\Require{$\rho \geq 1$} 
\Ensure{$X_k$} 
\While{$\left|J^{(t)}-J^{(t-1)}\right|<\epsilon$} 

\For{$k=1,...,K$}

\State{Solve $\nabla J_{k}\left(X_{i}^{k};U_{k},V_{k}\right)=\left[\frac{\partial J_{k}}{\partial\Gamma_{k}},\frac{\partial J_{k}}{\partial\mathbf{N}_{k}}\right]$} 

\State{$\hat{\beta}_{k}^{t}=\frac{1}{2}\cdot\frac{Tr\left[\Delta g(\mathbf{\Gamma}_{k})\cdot\Delta\mathbf{\Gamma}_{k}^{H}+\Delta\mathbf{\Gamma}_{k}\cdot\Delta g(\mathbf{\Gamma}_{k})^{H}\right]}{Tr\left[\Delta g(\mathbf{\Gamma}_{k})\cdot\Delta g(\mathbf{\Gamma}_{k})^{H}\right]}$}

\State{$\hat{\kappa}_{k}^{t}=\frac{1}{2}\cdot\frac{Tr\left[\Delta g(\mathbf{N}_{k})\cdot\Delta\mathbf{N}_{k}^{H}+\Delta\mathbf{N}_{k}\cdot\Delta g(\mathbf{N}_{k})^{H}\right]}{Tr\left[\Delta g(\mathbf{N}_{k})\cdot\Delta g(\mathbf{N}_{k})^{H}\right]}$}

\State{$\mathbf{\Gamma}^{(t+1)}_k=\mathbf{\Gamma}^{(t)}_k-\beta^{(t)}_k\frac{\partial J_k}{\partial\mathbf{\Gamma}_k}$} 

\State{$\mathbf{N}^{(t+1)}=\mathbf{N}^{(t)}-\kappa^{(t)}_k\frac{\partial J_k}{\partial\mathbf{N}_{k}}$}

\EndFor

\For{$k=1,...,K$}
\State{$w_{k}=\frac{\left(1/I_{k}\right)^{1/(p-1)}}{\sum_{k=1}^{K}\left(1/I_{k}\right)^{1/(p-1)}}$} 

\State{$J_{k}=w_{k}I_{k}$} 

\State{+= $\mu\sum_{q=1,q\neq k}^{K}\sum_{i.j}\left(d_{\mathbf{U}_{k},\mathbf{V}_{k}}(X_{i}^{k},X_{j}^{k})-d_{\mathbf{U}_{l},\mathbf{V}_{l}}(X_{i}^{q},X_{j}^{q})\right)^{2}$} 

\State{$J^{(t)} = J^{(t)} + J_{k}$}
\EndFor

\EndWhile 

\Return{$\mathbf{U}_{k}=\mathbf{\Gamma}_{k}^{T}\mathbf{\Gamma}_{k}$  and $\mathbf{V}_{k}=\mathbf{N}_{k}^{T}\mathbf{N}_{k}$}
\end{algorithmic}
\end{algorithm}

\section{Implementation\label{sec:Implementation}}
We develop a supporting functional library in Java (java-jdk/11.0.1),
and rely on a number of additional publicly available scientific
and mathematical open source packages including the Apache foundation
commons packages (e.g. Math Commons \citealp{Foundation2018a} and
Commons Lang \citealp{Foundation2018b}) and the JSOFA package to support our designs. The
overall functionality is supported at a high level by the following
open source packages:
\begin{itemize}
\item Maven is used to manage dependencies, and produce executable functionality
from the project \citet{Foundation2018}
\item JUnit is used to support library unit test management \citep{Junit2018} 
\item slf4j is used as a logging frame work \citep{QualityOpenSoftware2017}
\item MatFileRW is used for I/O handling \citep{MatFileRW2018}
\end{itemize}
We recommend reviewing the vsa-parent .pom file included as part of the software
package for a more comprehensive review of the functional dependency. Versions are subject to upgrades as development proceeds
beyond this publication. Python developer should note that the library outlined here can be easily accessed using any number of Python-to-Java projects (Py4J and Jython for example).  

Execution of the code was performed on a number
of platforms including a personal laptop (MacBook Pro, 2.5GHz Intel
Core i7, macOS Mojave) and an institution high performance computer
(Florida Institute of Technology, BlueShark HPC)\footnote{https://it.fit.edu/information-and-policies/computing/blueshark-supercomputer-hpc/}.
The development of the library and functionality in Java allow for
the functionality presented here to be applied regardless of platform.

We are not reporting processing times as part of this analysis as the computational times
varied depending on platform used. Our initial research
included using the parallel computing functionality packaged with
Java, in combination with the GPU functionality on the BlueShark computer.
Further research is necessary to quantify optimal implementation with
respect to convergence speed and memory usage.

\subsection{Training Data}

Similar to \citet{johnston2017variable}, we use the University of California Riverside Time Series Classification Archive (UCR)
and the Lincoln Near-Earth Asteroid Research (LINEAR) dataset to demonstrate the performance of our feature space classifier.
The individual datasets are described as follows:
\begin{itemize}
\item \textbf{UCR}: We baseline the investigated classification
methodologies \citep{Keogh2011} using the UCR time domain datasets. The UCR time domain dataset STARLIGHT \citep{Protopapas2006} is derived from a set of Cepheid, RR Lyra, and Eclipsing Binary Stars. This time-domain dataset is phased (folded) via the primary
period and smoothed using the SUPER--SMOOTHER algorithm \citep{Reimann1994}
by the Protopapas study prior to being provided to the UCR database. Note that the sub-groups of each of the three classes are combined together in the
UCR data (i.e., RR Lyr (ab) + RR Lyr (c) = RR). 

\item \textbf{LINEAR}: The original database LINEAR is subsampled; we select time series data that has been verified and for which accurate photometric values are available \citep{Sesar2011,Palaversa2013}. This subsampled set is parsed into separate training and test
sets.
From the starting sample of 7,194 LINEAR variables, a clean sample
of 6,146 time series datasets and their associated photometric values is used. Stellar class-type is limited further
to the top five most populous classes: RR Lyr (ab), RR Lyr (c), Delta
Scuti / SX Phe, Contact Binaries and Algol-Like Stars with 2 Minima,
resulting in a set of 6,086 time series datasets. 
\end{itemize}

Training data subsets are generated as follows: UCR already defines
a training and test set, the LINEAR data is split into a training
and test set using a predefined algorithm (random assignment, of
nearly equal representation of classes in training and test). We used a method
of 5-fold cross-validation both datasets; the partitions
in 5-fold algorithm are populated via random assignment. For more
details on the datasets themselves, a baseline for performance, and
additional references, see \citet{johnston2017variable}. 

\subsection{Feature Optimization}

The time domain transformation we selected requires parameter optimization
(resolution, kernel size, etc.); each survey can potentially have
a slightly different optimal set of transformation parameters with
respect to the underlying survey parameters (e.g. sample rate, survey
frequency, number of breaks over all observations, etc.). While we
could include the parameters optimization in the cross-validation
process for the classifier, this will be highly computationally challenging,
specifically for classifier that require iterations, as we would be
handling an increasing number of permutations with each iteration,
over an unknown number of iteration. To address this problem, the
feature space is cross-validated on the training dataset, and k-NN
classification is used (assuming a fixed temporary $k$ value allows
little to no tuning) to estimate the misclassification error with
the proposed feature space parameters. The optimized features are used as givens for the cross-validation process in optimizing
the intended classifier. Likely some loss of performance will occur,
but considering how the final classifier design is based on k-NN as
well, it is expected to be minor.

\begin{figure*}
\begin{centering}
\begin{tabular}{cc}
\includegraphics[scale=0.2]{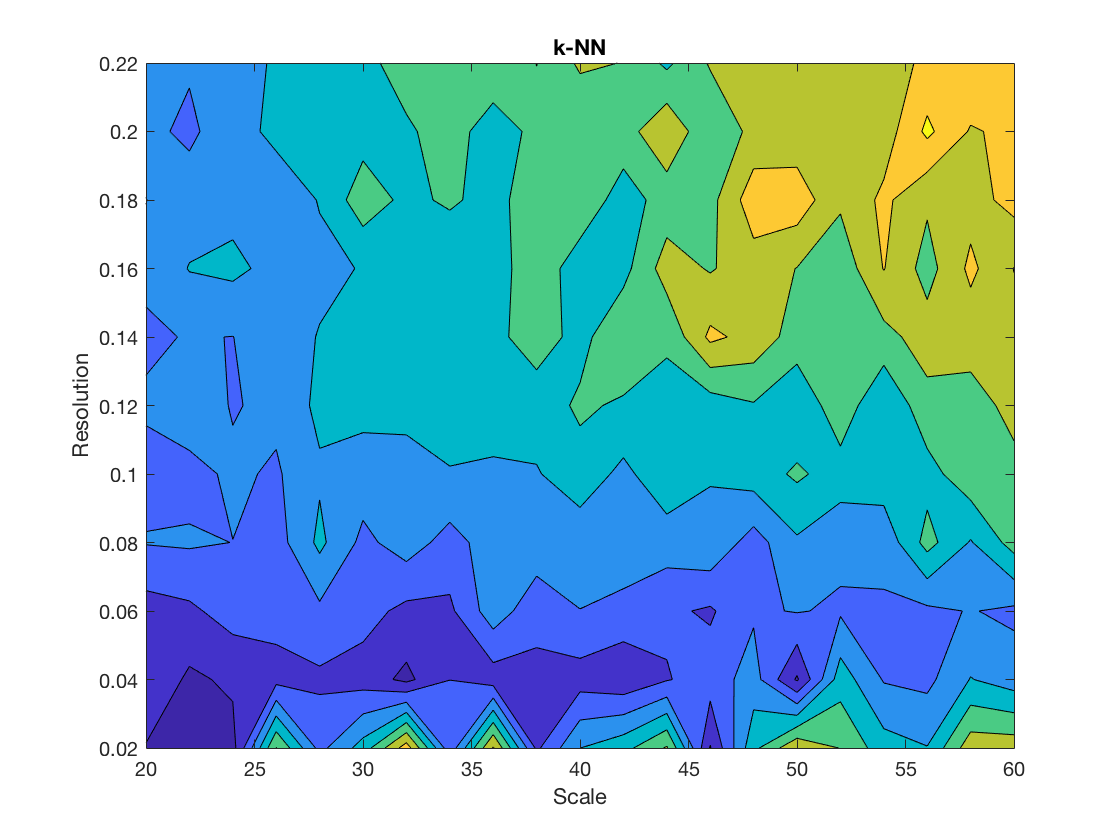} & \includegraphics[scale=0.2]{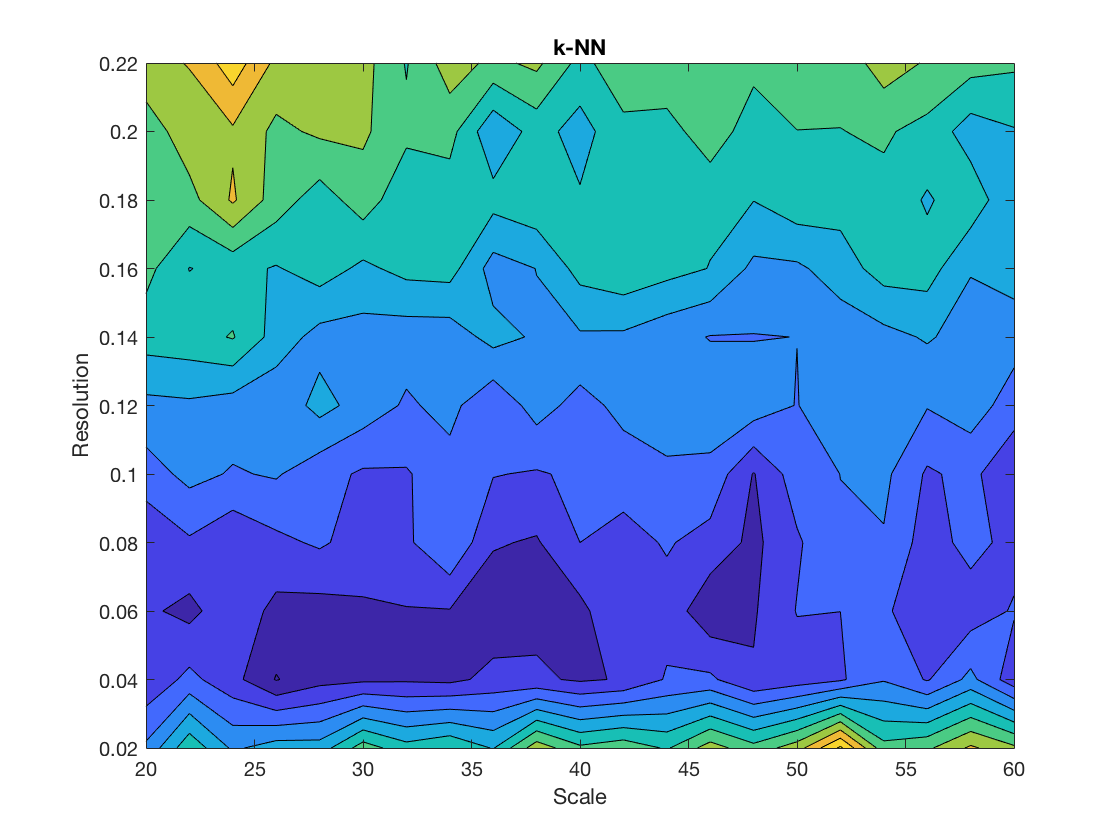}\tabularnewline
\end{tabular}
\par\end{centering}
\caption{Parameter optimization of the SSMM feature space (Left: UCR Data,
Right: LINEAR Data)\label{fig:Optimization-of-SSMM}}
\end{figure*}

Because of the multi-dimensional nature of our feature space, we propose the following method for feature optimization---per class we generate
a mean representation of the feature (given the fraction of data being
trained on), all data are then transformed (training and cross-validation
data) via \citet{park2003lower} into a distance representation, i.e.
the difference of the observed feature and each of the means is generated.
Note that for the matrix feature spaces, the Frobenius Norm is used.
Alternatively we could have generated means based on unsupervised clustering
(k-Means); while not used in this study, this functionality is provided as part of the code. We found that the performance using the unsupervised case was
very sensitive to the initial number of $k$ used. For the LINEAR
and UCR datasets, the results were found with respect to optimization
of feature (DF and SSMM) parameters to be roughly the same. A k-NN
algorithm is applied to the reduced feature space, 5-fold classification
is then used to generate the estimate of error, and the misclassification
results are presented a response map given feature parameters (Figures
\ref{fig:Optimization-of-Distribution} and Figure \ref{fig:Optimization-of-SSMM}):

\begin{figure*}
\begin{centering}
\begin{tabular}{cc}
\includegraphics[scale=0.2]{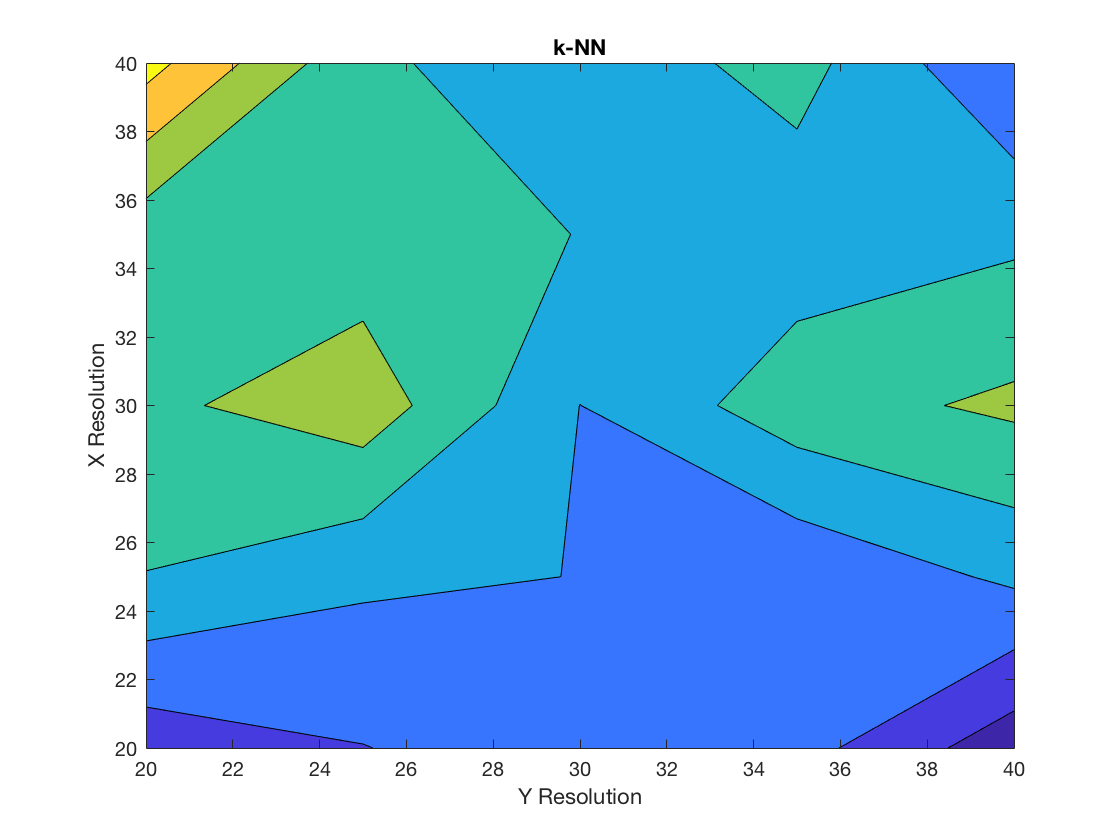} & \includegraphics[scale=0.2]{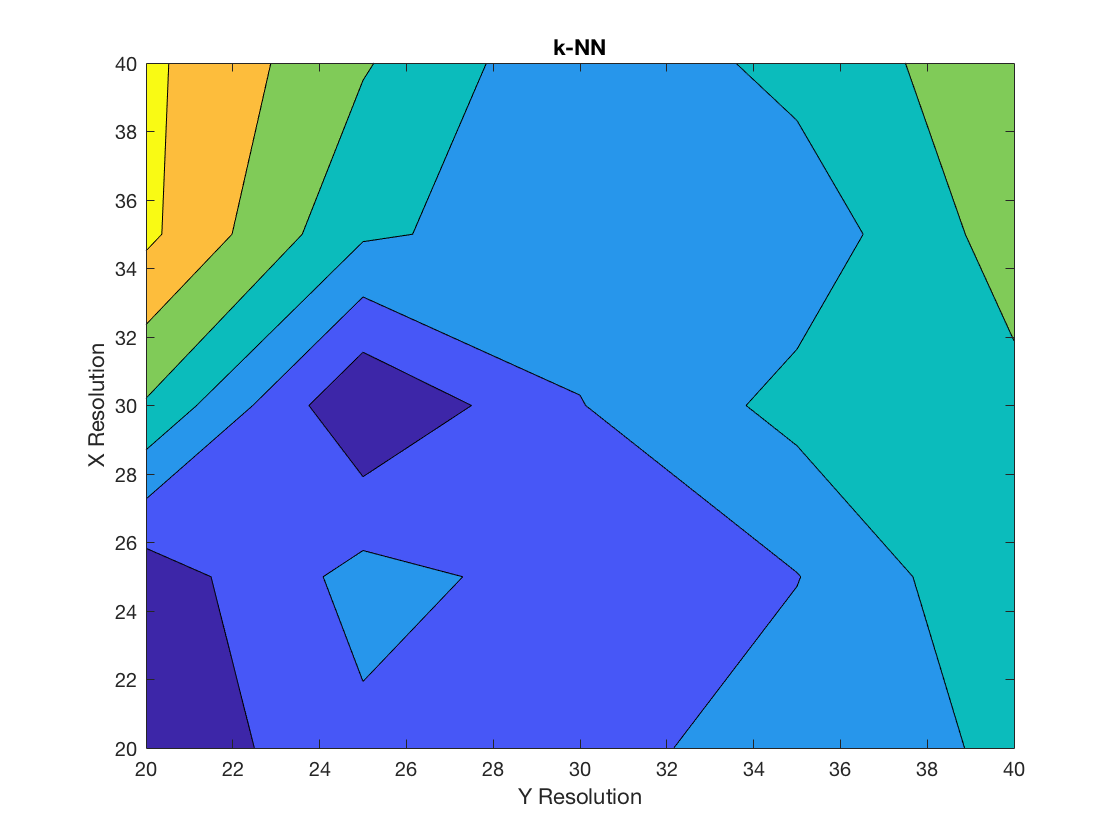}\tabularnewline
\end{tabular}
\par\end{centering}
\caption{Parameter optimization of the Distribution Field feature space (Left:
UCR Data, Right: LINEAR Data)\label{fig:Optimization-of-Distribution}}
\end{figure*}

We select the optimum values for each feature space, based on a
minimization of both the LINEAR and UCR data. These values are estimated
to be: DF Optimized (x, y) -- $30\times25$, SSMM Optimized (res x scale) --
$0.06\times35.0$ 

\subsection{Large Margin Multi-view Metric Learning Implementation}

The implementation of LM\textsuperscript{3}L is applied to the UCR and LINEAR
datasets. Based on the number of views associated with each feature
set, the underlying classifier will be different (e.g. UCR does not
contain color information and it also has only three classes compared
to LINEAR's five). The features SSMM, DF, and Statistical
Representations (mean, standard deviation, kurtosis, etc.) are computed
for both datasets. Color and the time domain representations provided
with the LINEAR data are also included as additional views. 

\begin{figure}
\centering{}\includegraphics[scale=0.2]{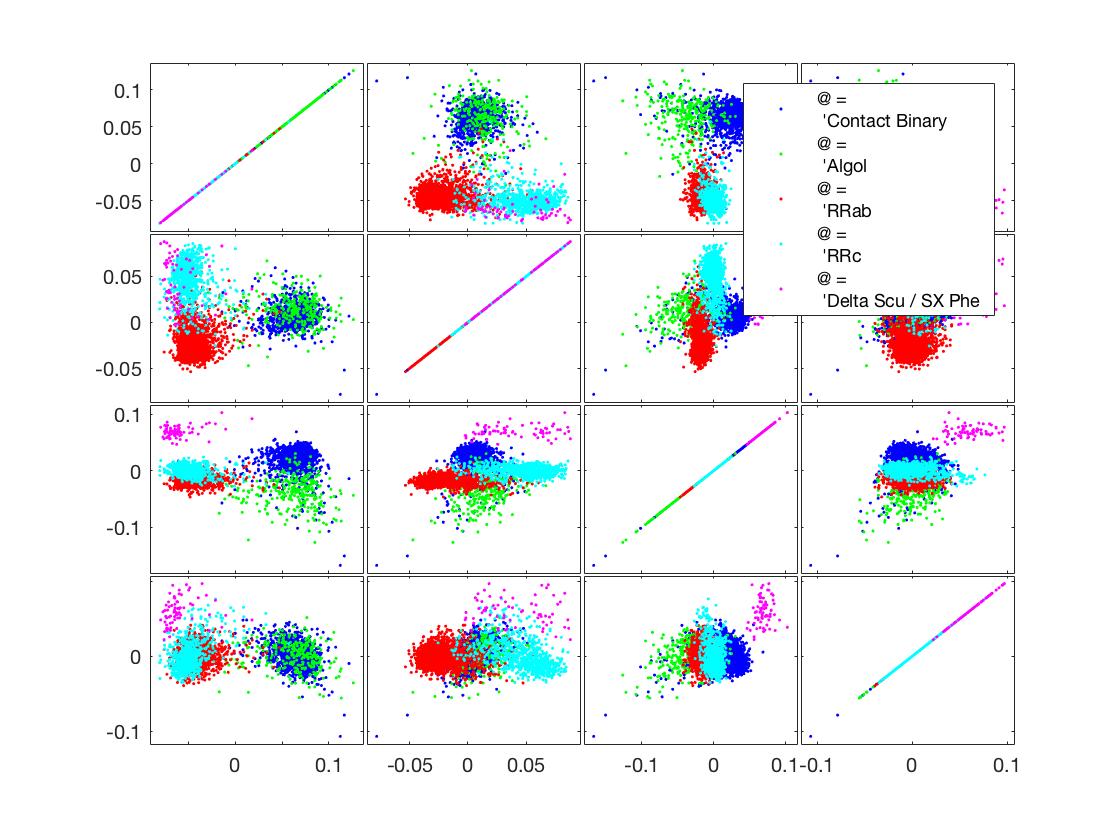}
\caption{DF Feature space after ECVA reduction from LINEAR (Contact Binary/blue
circle, Algol/ red +, RR (ab)/green points, RR (c) in black squares, Delta
Scu/SX Phe magneta diamonds) off-diagonal plots represent comparison
between two different features, on-diagonal plots represent distribution
of classes within a feature (one dimensional) \label{fig:DF-Feature-Space}}
\end{figure}

\begin{figure}
\centering{}\includegraphics[scale=0.2]{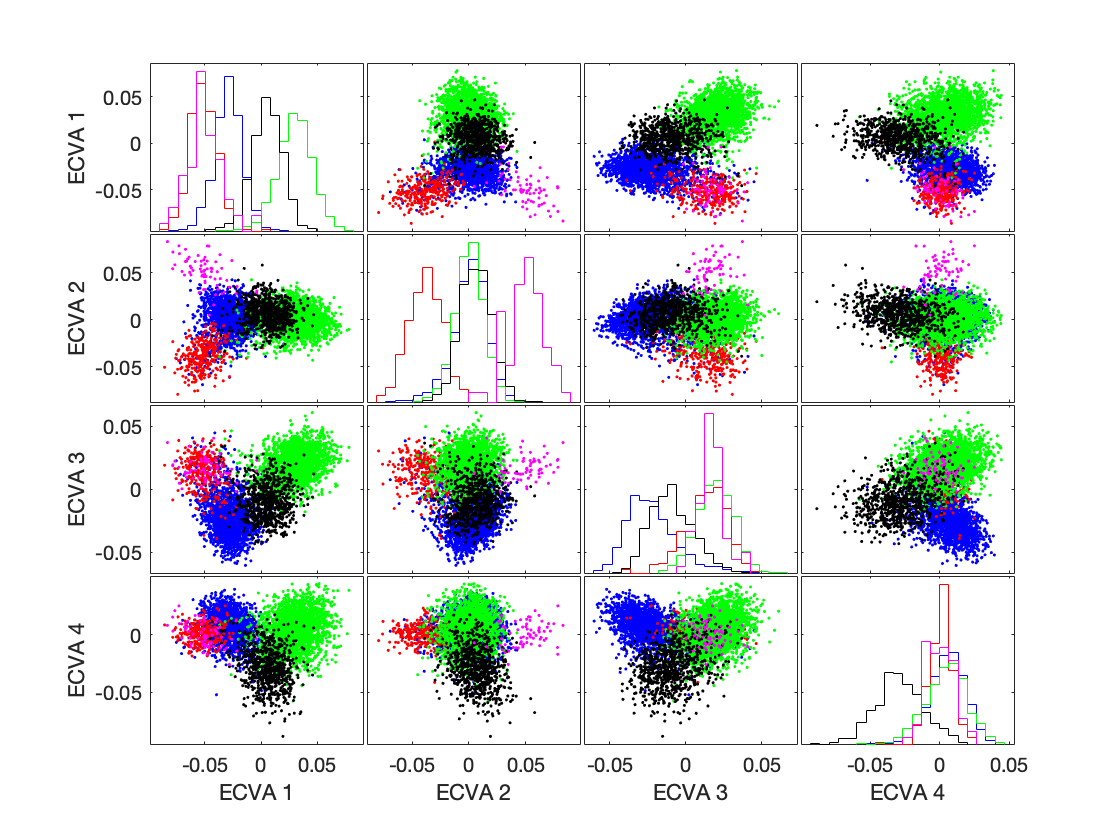}
\caption{SSMM feature space after ECVA reduction LINEAR (Contact Binary/blue
circle, Algol/ red +, RR (ab)/green points, RR (c) in black squares, Delta
Scu/SX Phe magneta diamonds) off-diagonal plots represent comparison
between two different features, on-diagonal plots represent distribution
of classes within a feature (one dimensional). The dimension plotted are the first 4 canonical variates (transformed DF space).  \label{fig:SSMM-Feature-Space}}
\end{figure}

To allow for the implementation of the vector-variate classifier,
the dimensionality of the SSMM and DF features are reduced via vectorization
of the matrix and then processing by the ECVA algorithm, resulting
in a dimensionality that is $k-1$, where $k$ is the number of classes. We note that without this processing via ECVA, the times for
the optimization became prohibitively long, this is similar to the implementation of IPMML given in \cite{zhou2016}. SSMM and DF features are
generated with respect to the LINEAR dataset---Park's transformation
is not applied here---the feature space reduced via ECVA, the
results are and given in Figure \ref{fig:DF-Feature-Space} (DF) and
Figure \ref{fig:SSMM-Feature-Space} (SSMM). 

\begin{figure*}
\centering{}%
\begin{tabular}{cc}
\includegraphics[scale=0.2]{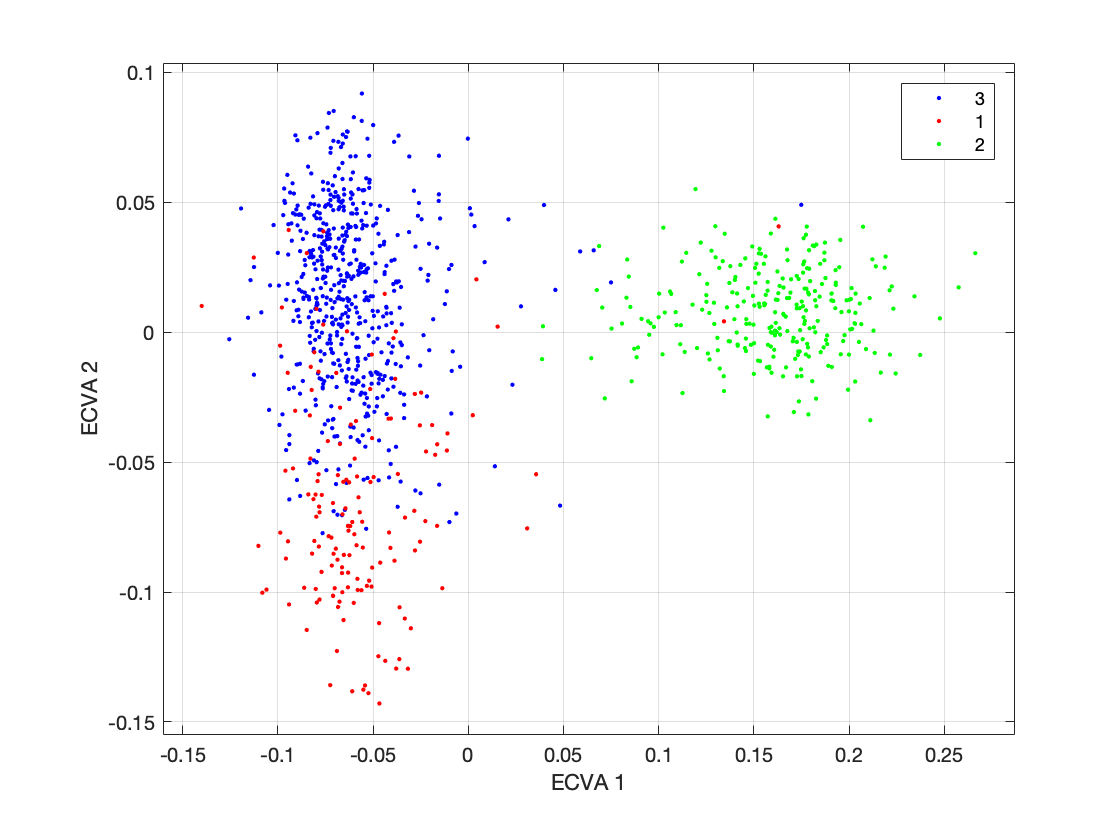} & \includegraphics[scale=0.2]{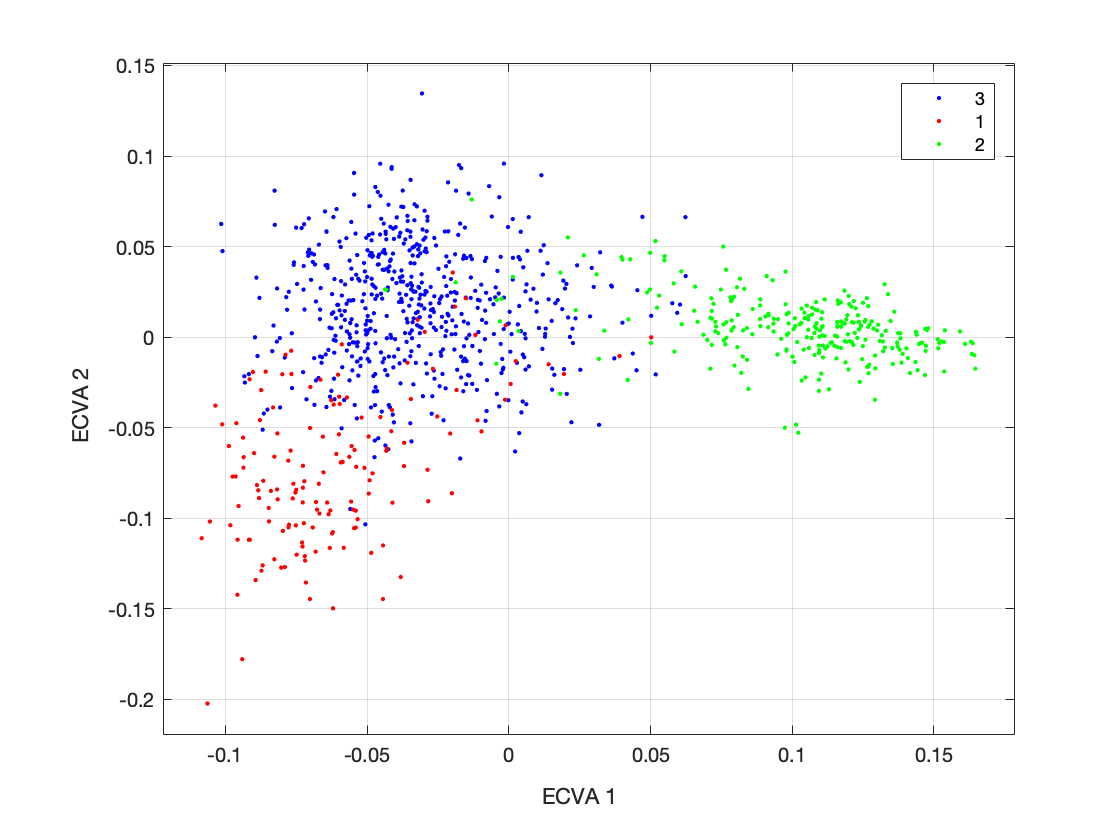}\tabularnewline
\end{tabular}
\caption{DF--Left and SSMM--Right feature space after ECVA reduction from
UCR. Class names (1,2, and 3) are based on the classes provided by
the originating source and the UCR database. The dimension plotted are the first 2 canonical variates (transformed DF space). \label{fig:DF-Feature-Space-1}}
\end{figure*}

Similarly, the SSMM and DF features are generated for the UCR dataset---Park's transformation is not applied here--and the feature space
reduced via ECVA. The results are plotted and given in Figure \ref{fig:DF-Feature-Space-1}
(DF--Left) and (SSMM--Right). 

The dimensions given in the figures are reduced dimensions resulting
from the ECVA transform and therefore they do not necessarily have
meaningful descriptions (besides $x_{1},x_{2},x_{3},...,x_{n}$).
These reduced feature spaces are used as input to the LM\textsuperscript{3}L algorithm.

The individual views are standardized (subtract by mean and divide
by standard deviation). Cross-validation of LM\textsuperscript{3}L is 
used to optimize the three tunable parameters and the one parameter
associated with the k-NN. The LM\textsuperscript{3}L authors recommend
some basic parameter starting points; our analysis includes investigating
the tunable values as well as an upper (+1) and lower (-1) level 
about each parameter, over a set of odd k-NN values {[}1,19{]}; 
the optimization only needs to occur for each set of tunable values, 
the misclassification given a k-Value can be evaluated separately, 
this experiment is outlined in Table \ref{tab:The-Cross-Validation-Process}. 

\begin{table}

\caption{The cross-validation process for LM\textsuperscript{3}L tunable values for each of the seven experimental settings. $\tau$ and $\eta$ define the large margin threshold, $\mu$ controls the importance of pairwise distance between views  \label{tab:The-Cross-Validation-Process}}
\centering{}%
\begin{tabular}{|c|c|c|c|c|c|c|c|}
\hline 
Variable & 1 & 2 & 3 & 4 & 5 & 6 & 7\tabularnewline
\hline 
\hline 
$\tau$ & 1 & 1.75 & 0.25 & 1 & 1 & 1 & 1\tabularnewline
\hline 
$\eta$ & 5 & 5 & 5 & 8 & 2 & 5 & 5\tabularnewline
\hline 
$\mu$ & 0.5 & 0.5 & 0.5 & 0.5 & 0.5 & 1.0 & 0.1\tabularnewline
\hline 
\end{tabular}
\end{table}

Cross-validation is performed to both optimize for our application
and investigate the sensitivity of the classifier to adjustment of
these parameters. For a break down of the cross validation results,
see the associated datasets and spreadsheet provided as part of the
digital supplement.

\subsubsection{Testing and Results (UCR \& LINEAR)}

Based on the cross-validation process, the following optimal parameters
are found:
\begin{itemize}
\item {\it LINEAR}: k-NN(11), $\tau$(1.0), $\eta$(5.0), $\mu$(0.1) 
\item {\it UCR}: k-NN(9), $\tau$(1.0), $\eta$(5.0), $\mu$(0.1) 
\end{itemize}

The classifier is then trained using the total set of training data
along with the optimal parameters selected. Given
\cite{hu2014large} definitions: $\mu$ controls the importance 
of pairwise distance between views in the optimization process,
$\tau$ and $\eta$ define the threshold $y_{ij}(\eta_k-d^2_{M_k}(x^k_i,x^k_j))>\tau_k$
used as the margin definition. The trained classifier is applied to the 
test data, the confusion matrices \citep{fawcett2006introduction} resulting 
from the application are presented in Table \ref{tab:LINEAR-Confusion-Matrix}
and Table \ref{tab:UCR-Confusion-Matrix}:

\begin{table*}
\caption{LINEAR confusion matrix via LM\textsuperscript{3}L entries are counts (percent)
\label{tab:LINEAR-Confusion-Matrix}}
\begin{centering}
\begin{tabular}{|c|c|c|c|c|c||c|}
\hline 
Misclassification Rate & RR Lyr (ab) & Delta Scu / SX Phe & Algol & RR Lyr (c) & Contact Binary & Missed\tabularnewline
\hline 
\hline 
RR Lyr (ab) & 1081 (0.992) & 0 (0.000) & 0 (0.000) & 6 (0.006) & 1 (0.001) & 2 (0.002)\tabularnewline
\hline 
Delta Scu / SX Phe & 0 (0.000) & 23 (0.852) & 0 (0.000) & 2 (0.074) & 2 (0.074) & 0 (0.000)\tabularnewline
\hline 
Algol & 1 (0.007) & 0 (0.000) & 108 (0.788) & 0 (0.000) & 28 (0.204) & 0 (0.000)\tabularnewline
\hline 
RR Lyr (c) & 23 (0.062) & 0 (0.000) & 1 (0.003) & 343 (0.925) & 4 (0.011) & 0 (0.000)\tabularnewline
\hline 
Contact Binary & 3 (0.003) & 0 (0.000) & 29 (0.033) & 9 (0.010) & 832 (0.952) & 1 (0.001)\tabularnewline
\hline 
\end{tabular}
\par\end{centering}
\end{table*}

\begin{table*}
\centering{}\caption{UCR confusion matrix via LM\textsuperscript{3}L entries are counts (percent)\label{tab:UCR-Confusion-Matrix}}
\begin{tabular}{|c|c|c|c||c|}
\hline 
Misclassification Rate & 2 & 3 & 1 & Missed\tabularnewline
\hline 
\hline 
2 & 2296 (0.996) & 9 (0.004) & 0 (0.000) & 0 (0.000)\tabularnewline
\hline 
3 & 17 (0.004) & 4621 (0.972) & 116 (0.023) & 0 (0.000)\tabularnewline
\hline 
1 & 8 (0.007) & 375 (0.319) & 794 (0.675) & 0 (0.000)\tabularnewline
\hline 
\end{tabular} 
\end{table*}

\subsection{Large Margin Multi-View Metric Learning with Matrix Variates Implementation}

The implementation of LM\textsuperscript{3}L-MV is applied to the UCR and LINEAR
datasets. The features SSMM, DF, and Statistical
Representations (mean, standard deviation, kurtosis, etc.) are computed
for both datasets. Similar to the LM\textsuperscript{3}L
procedure, color and the time domain representations provided with
the LINEAR data are also included as additional views. The implementation
of the matrix-variate classifier, allows us to avoid the vectorization
and feature reduction (ECVA) step. The individual views are standardized
prior to optimization. Also similar to the LM\textsuperscript{3}L procedure, cross-validation
of LM\textsuperscript{3}L-MV is used to optimize the three tunable parameters
and the one parameter associated with the k-NN. The table of explored
tunable parameters is given in Table \ref{tab:The-Cross-Validation-Process-1}:

\begin{table*}
\caption{The cross-validation process for LM\textsuperscript{3}L-MV tunable values \label{tab:The-Cross-Validation-Process-1}}
\begin{centering}
\begin{tabular}{|c|c|c|c|c|c|c|c|}
\hline 
Variable & 1 & 2 & 3 & 4 & 5 & 6 & 7\tabularnewline
\hline 
\hline 
$\lambda$ & 0.5 & 1.0 & 0.25 & 0.5 & 0.5 & 0.5 & 0.5\tabularnewline
\hline 
$\mu$ & 0.5 & 0.5 & 0.5 & 1.0 & 0.25 & 0.5 & 0.5\tabularnewline
\hline 
$\gamma$ & 0.5 & 0.5 & 0.5 & 0.5 & 0.5 & 1.0 & 0.25\tabularnewline
\hline 
\end{tabular}
\par\end{centering}
\end{table*}

For a break down of the results, see the associated datasets and spreadsheet
provided as part of the digital supplement.

\subsubsection{Testing and Results (UCR \& LINEAR)}

Based on the cross-validation process, the following optimal parameters
are found (and their cross-validation error estimates):
\begin{itemize}
\item {\it LINEAR}: k-NN(15), $\lambda$(0.5), $\mu$(0.5), $\gamma$(0.5) 
\item {\it UCR}: k-NN(19), $\lambda$(0.5), $\mu$(1.0), $\gamma$(0.5)
\end{itemize}
The classifier is then trained using the total set of training data
along with the optimal parameters selected. The $\lambda$ parameter controls the importance of regularization, the $\mu$ parameter controls the importance of pairwise distance in the optimization process, and $\gamma$ controls the balance between push and pull.

The trained classifier
is applied to the test data, the confusion matrices resulting from
the application are presented in Table \ref{tab:LINEAR-Confusion-Matrix-1}
and Table \ref{tab:UCR-Confusion-Matrix-1}:

\begin{table*}
\caption{LINEAR confusion matrix via LM\textsuperscript{3}L-MV entries are counts (percent)
\label{tab:LINEAR-Confusion-Matrix-1}}
\centering{}%
\begin{tabular}{|c|c|c|c|c|c||c|}
\hline 
Misclassification Rate & RR Lyr (ab) & Delta Scu / SX Phe & Algol & RR Lyr (c) & Contact Binary & Missed\tabularnewline
\hline 
\hline 
RR Lyr (ab) & 1074 (0.985) & 0 (0.000) & 1 (0.001) & 15 (0.014) & 0 (0.000) & 0 (0.000)\tabularnewline
\hline 
Delta Scu / SX Phe & 1 (0.037) & 24 (0.889) & 0 (0.000) & 2 (0.074) & 0 (0.000) & 0 (0.000)\tabularnewline
\hline 
Algol & 3 (0.022) & 0 (0.000) & 104 (0.759) & 1 (0.007) & 29 (0.212) & 0 (0.000)\tabularnewline
\hline 
RR Lyr (c) & 23 (0.059) & 0 (0.000) & 1 (0.003) & 343 (0.930) & 4 (0.008) & 0 (0.000)\tabularnewline
\hline 
Contact Binary & 3 (0.003) & 0 (0.000) & 29 (0.035) & 9 (0.001) & 832 (0.958) & 1 (0.002)\tabularnewline
\hline 
\end{tabular}
\end{table*}

\begin{table*}
\centering{}\caption{UCR confusion matrix via LM\textsuperscript{3}L-MV entries are counts (percent)
\label{tab:UCR-Confusion-Matrix-1}}
\begin{tabular}{|c|c|c|c||c|}
\hline 
Misclassification Rate & 2 & 3 & 1 & Missed\tabularnewline
\hline 
\hline 
2 & 2298 (0.997) & 6 (0.003) & 0 (0.000) & 1 (\textasciitilde{}0.000)\tabularnewline
\hline 
3 & 4 (0.001) & 4450 (0.936) & 300 (0.063) & 0 (0.000)\tabularnewline
\hline 
1 & 3 (0.003) & 467 (0.397) & 707 (0.601) & 0 (0.000)\tabularnewline
\hline 
\end{tabular} 
\end{table*}

\subsection{Comparison}

The matrix-variate and the vector-variate versions do not perform
much different under the conditions provided given the data observed.
However, as a reminder, the LM\textsuperscript{3}L implementation includes a feature
reduction methodology (ECVA) that our LM\textsuperscript{3}L-MV does not. The
ECVA front end was necessary as the dimensionality of the unreduced
input vectors results in features and metrics which are prohibitively
large (computationally).
It is not entirely surprising that our two
competitive methodologies perform similarly (see Tables \ref{tab:LINEAR-Confusion-Matrix-Delta} and \ref{tab:UCR-Confusion-Matrix-Delta}), with the LM\textsuperscript{3}L algorithm
of having the benefit of being able to process the matrix-variate
spaces ahead of time via ECVA and thus being able to process the SSMM
and DF features spaces in a lower dimension ($c-1$ dimensions). 
For a quantitative comparison of our classifiers, we have computed precision
and recall metrics for our presented classifiers \citep{fawcett2006introduction,sokolova2009systematic} as
well as an overall estimate of F1-score, the results of this analysis
are presented in Table \ref{tab:Performance-Metrics-for}. However, a direct one-to-one comparison to other pattern classification methods is difficult, as there are no other classifiers that we know of that are both multi-view and matrix-variate. 

We can still provide some context by looking at known alternatives that may partially address our particular situation. We have included results based on the implementation of a multi-view k-NN classifier (\ref{eq:Distance}) in both the matrix-variate and vector-variate domains, but with the metrics being the identity matrix (i.e. Euclidean and Forbinus distances respectively), as a baseline reference point. Similarly, we have included results based on the implementation of a single view k-NN classifier (\ref{eq:Distance}) applied to the vectorized and concatenated features (i.e. Euclidean distance).

For comparison we include classifiers generated from the main individual feature spaces; optimization was performed using Random Forest classification \citep{Breiman1984}. In addition to these standard methods applied to the unreduced feature space/views, we have generated classifiers based on the dimensionally reduced feature space generated resulting from the ECVA algorithm applied to the DF and SSMM vectorized feature spaces. These reduced feature spaces/views are implemented using the \cite{zhou2016} IPMML (i.e., multi-view algorithm), this implementation is the nearest similar implementation to both our LM\textsuperscript{3}L and LM\textsuperscript{3}L-MV algorithm designs.  Detailed computations associated with all analyses are included as
part of the digital supplement.

\begin{table}

\caption{F1-Score metrics for the proposed classifiers with respect to LINEAR
and UCR datasets \label{tab:Performance-Metrics-for}}
\begin{centering}
\begin{tabular}{|c|c|c|}
\hline 
F1-Score & UCR & LINEAR\tabularnewline
\hline 
\hline 
LM\textsuperscript{3}L & 0.904 & 0.918\tabularnewline
\hline 
LM\textsuperscript{3}L-MV & 0.860 & 0.916\tabularnewline
\hline 
IPMML & 0.900 & 0.916\tabularnewline
\hline 
k-NN Multi-View MV & 0.725 & 0.574\tabularnewline
\hline 
k-NN Multi-View & 0.691 & 0.506\tabularnewline
\hline 
k-NN Concatenated & 0.650 & 0.427\tabularnewline
\hline 
RF -- DF & 0.878 & 0.650\tabularnewline
\hline 
RF -- SSMM & 0.659 & 0.402\tabularnewline
\hline 
RF -- Time Statistics & 0.678 & 0.787\tabularnewline
\hline 
\end{tabular}
\par\end{centering}
\end{table}

It should be noted, that ECVA has its limitations; anecdotally on
more then one occasion during the initial analysis, when the full
dataset was provided to the algorithm, the memory of the machine was
exceeded. Care was taken with the LINEAR dataset to develop a training
dataset that was small enough that the out-of-memory error would not
occur, but a large enough that each of the class-types was represented
sufficiently. Similarly, the projection into lower dimensional space
meant that the LM\textsuperscript{3}L implementation iterated at a much faster
rate with the same amount of data, compared to the LM\textsuperscript{3}L-MV algorithm.
The matrix multiplication operations associated with the matrix distance
computation are more computationally expensive compared to the simpler
vector metric distance computation, however many computational languages have been optimized for matrix multiplication (e.g., MATLAB, Mathmatica, CUDA, etc.). Again, the time the ECVA algorithm
takes to operate upfront saves the LM\textsuperscript{3}L iterations time. In
general, both algorithms perform well with respect to misclassification
rate, but both also require concessions to handle the scale and scope
of the feature spaces used. The cost of most of these concession can
be mitigated with additional machine learning strategies, some of
which we have begun to implement here---parallel computation for example.

\section{Conclusions}

The classification of variable stars relies on a proper selection
of features of interest and a classification framework that can support
the linear separation of those features. Features should be selected
that quantify the signature of the variability, i.e. its
structure and information content. Here, two features which have utility
in providing discriminatory capabilities, the SSMM and DF feature
spaces are studied. The feature extraction methodologies are applied
to the LINEAR and UCR dataset, as well as a standard breakdown of
time domain descriptive statistics, and in the case of the LINEAR
dataset, a combination of $ugriz$ colors. To support the set of high-dimensionality
features, or views, multi-view metric learning is investigated as
a viable design. Multi-view learning provides an avenue for integrating
multiple transforms to generate a superior classifier. The structure
of multi-view metric learning allows for a number of modern computational
designs to be used to support increasing scale and scope (e.g., parallel
computation); these considerations can be leveraged given the parameters
of the experiment designed or the project in question. 

Presented, in addition to an implementation of a standard multi-view
metric learning algorithm (LM\textsuperscript{3}L) that works with a feature space
that has been vectorized and reduced in dimension, is a multi-view
metric learning algorithm designed to work with matrix-variate views.
This new classifier design does not require transformation of the
matrix-variate views ahead of time, and instead operates directly
on the matrix data. The development of both algorithm designs (matrix-variate
and vector-variate) with respect to the targeted experiment of interest
(discrimination of time-domain variable stars) highlighted a number
of challenges to be addressed prior to practical application. In overcoming
these challenges, it was found that the novel classifier design (LM\textsuperscript{3}L-MV)
performed on order of the staged (Vectorization + ECVA + LM\textsuperscript{3}L)
classifier. Future research will include investigating overcoming
high dimensionality matrix data (e.g. SSMM), improving the parallelization
of the design presented, and implementing community standard workarounds for large dataset data (i.e., on-line learning, stochastic/batch
gradient descent methods, k-d tree... etc.).

\section*{Acknowledgements}

The authors are grateful for the valuable machine learning discussion
with S. Wiechecki Vergara. Inspiration provided by C. Role. Research was partially supported
by Perspecta, Inc. This material is based upon work supported by the
NASA Florida Space Grant under 2018 Dissertation And Thesis Improvement
Fellowship (No. 202379). The LINEAR program is sponsored by the National
Aeronautics and Space Administration (NRA Nos. NNH09ZDA001N, 09-NEOO09-0010)
and the United States Air Force under Air Force Contract FA8721-05-C-0002.
This material is based upon work supported by the National Science
Foundation under Grant No. CNS 09-23050.

\bibliographystyle{plainnat}
\bibliography{L3ML}

\appendix 

\section{Challenges Addressed}
\label{sec:challenges}
In the application of the LM\textsuperscript{3}L algorithm to our data we found
a number of challenges not specified by the original paper that required
attention. Some of these challenges were a direct result of our views
(vectorization of matrix-variate data) and some of these challenges
were resulting from practical application (hinge loss functionality
and step-size optimization).

\subsection{Hinge Loss Functionality\label{subsec:Hinge-Loss-Functionality}}

While the original LM\textsuperscript{3}L paper does not specify details with
respect to the implementation of the hinge loss functionality used,
the numerical implementation of both the maxima and the derivative
of the maxima are of critical importance. For the implementation here,
the hinge-loss functionality is approximated using Generalized Logistic
Regression \citep{zhang2001text,rennie2005loss}. Should a different
approximation of hinge loss be requested, care should be given to
the implementation, as definitions from various public sources are
not consistent. For purposes here, the Generalized Logistic Regression
is used to approximate the hinge loss ($h[x]\approx g_{+}\left(z,\phi\right)$)
and is defined as equation \ref{eq:13}:

\begin{equation}
g_{+}(z,\phi)=\left\{ \begin{array}{cc}
0.0 & z\leq-10\\
z & z\geq10\\
\frac{1}{\phi}\log\left(1+\exp\left(z\phi\right)\right) & -10<z<10
\end{array}\right.\label{eq:13}
\end{equation}

the derivative of the Generalized Logistic Regression is then given
as equation \ref{eq:14}:

\begin{equation}
\frac{\partial g_{+}\left(z,\phi\right)}{\partial z}=\left\{ \begin{array}{cc}
0.0 & z\leq-10\\
1 & z\geq10\\
\frac{\exp\left(\phi z\right)}{1+\exp\left(\phi z\right)} & -10<z<10
\end{array}\right.\label{eq:14}
\end{equation}

For practical reasons (underflow/overflow) the algorithm is presented
as a piece-wise function, in particular this is necessary because
of the exponential in the functionality. In addition, the public literature
is not consistent on the definition of the hinge-loss functionality
approximation, specifically the relationship between the notations:
$\left[z\right]_{+}$, $h[z]$, $\max(z,0)$, and $g_{+}\left(z,\phi\right)$;
usually the inconsistency is with respect to the input i.e. $z$,
$-z$, or $1-z$. We have explicitly stated our implementation here
to eliminate any confusion. 

\subsection{Step-Size Optimization}

While LM\textsuperscript{3}L provides an approximate "good" step size to use,
in practice we found that a singular number was not necessarily
useful. While the exact reasons of why a constant step size was not
beneficial were not investigated; the following challenges were identified: 
\begin{enumerate}
\item The possibility of convergence was very sensitive to the step size.
\item Small step sizes that did result in a consistent optimization, resulted
in a very slow convergence.
\item While an attempt could be made to find an optimal step size with respect
to all views, it seems unlikely this would occur given the disparate
nature of the views we have selected (distribution field, photometric
color, time domain statistics, etc.).
\item For the metric learning methods used here (in both the standard and
the proposed algorithms) the objective function magnitude scales with
the number of training data sets, view dimensions and the number of
views, as is apparent from the individual component of $LM^{3}L$:
$\sum_{i,j,}h\left[\tau_{k}-y_{ij}\left(\eta_{k}-d_{\mathbf{M}_{k}}^{2}(x_{i}^{k},x_{j}^{k})\right)\right]$.
With increasing number of training data, the objective function will
increase and the gradient component ($w_{k}^{p}\sum_{i,j}y_{ij}h'[z]\mathbf{C}_{ij}^{k}$)
will similarly be effected. This means that computational overflows
could occur just by increasing the number of training data used.
\end{enumerate}
In lieu of a singular estimate, we propose a dynamic estimate of the
step-size per iteration per view. A review of step-size and gradient
descent optimization methods \citep{ruder2016overview} suggest a
number of out-of-the-box solutions to the question of speed (specifically
methods such as Mini-Batch gradient descent). 

The question of dynamic step size requires more development, in particular
while methods exists, these are almost entirely focused on vector
variate optimization. \citet{barzilai1988two} outline a method for
dynamic step size estimation that has its' basis in secant root finding,
the method described is extended here to allow for matrix variate
cases. The gradient descent update for our metric learning algorithm
is given as equation \ref{eq:15}.

\begin{equation}
\mathbf{L}^{(t+1)}=\mathbf{L}^{(t)}-\beta\frac{\partial J}{\partial\mathbf{L}}\label{eq:15}
\end{equation}

In the spirit of Barziliai and Borwein, here in known as the BB-step
method, the descent algorithm is reformulated as equation \ref{eq:16}:

\begin{equation}
\lambda_{k}=\arg\min_{\lambda}\left\Vert \Delta\mathbf{L}-\lambda\Delta g(\mathbf{L})\right\Vert _{F}^{2}\label{eq:16}
\end{equation}

where $\lambda_{k}$ is a dynamic step size to be estimated per iteration
and per view, $\triangle g(\mathbf{L})=\nabla f\left(\mathbf{L}^{(t)}\right)-\nabla f\left(\mathbf{L}^{(t-1)}\right)$
and $\Delta\mathbf{L}=\mathbf{L}^{(t)}-\mathbf{L}^{(t-1)}$. The Forbinus
norm can be defined as $\left\Vert A\right\Vert _{F}^{2}=Tr(A\cdot A^{H})$,
the BB-step method can be found as equation \ref{eq:17}:

\begin{equation}
\frac{\partial}{\partial\lambda}Tr\left[\left(\Delta\mathbf{L}-\lambda\Delta g(\mathbf{L})\right)\left(\Delta\mathbf{L}-\lambda\Delta g(\mathbf{L})\right)^{H}\right]=0\label{eq:17}
\end{equation}

Based on the Matrix Cookbook \citep{petersen2008matrix}, equation
\ref{eq:17} can be transformed into equation \ref{eq:18}.

\begin{equation}
Tr\left[-\Delta g(L)\left[\Delta\mathbf{L}-\lambda\Delta g(\mathbf{L})\right]^{H}- \left[\Delta\mathbf{L}-\lambda\Delta g(\mathbf{L})\right]\Delta g(L)^{H}\right]=0\label{eq:18}
\end{equation}

With some algebra, equation \ref{eq:18} is turned into a solution
for an approximation of optimal step size, given here as equation
\ref{eq:19}.

\begin{equation}
\hat{\lambda}=\frac{1}{2}\cdot\frac{Tr\left[\Delta g(\mathbf{L})\cdot\Delta\mathbf{L}^{H}+\Delta\mathbf{L}\cdot\Delta g(\mathbf{L})^{H}\right]}{Tr\left[\Delta g(\mathbf{L})\cdot\Delta g(\mathbf{L})^{H}\right]}\label{eq:19}
\end{equation}

It is elementary to show that our methodology can be extended for
$\triangle g(\mathbf{L}_{k})=\nabla f\left(\mathbf{L}_{k}^{(t)}\right)-\nabla f\left(\mathbf{L}_{k}^{(t-1)}\right)$
and $\Delta\mathbf{L}_{k}=\mathbf{L}_{k}^{(t)}-\mathbf{L}_{k}^{(t-1)}$;
likewise we can estimate $\hat{\lambda}_{k}$ per view, so long as
the estimates of both gradient and objective function are monitored
at each iteration. While this addresses our observations, it should
be noted that the fourth challenged outlined (scaling with increasing
features and training data) was only partially addressed. Specifically,
the above methodology does not address an initial guess of $\lambda_{k}$;
in multiple cases it was found that this initial value was set to
high, causing our optimization to diverge. Providing an initial metric
in the form of $\sigma\mathbb{I}$ where $0<\sigma<1$ , was found
to improve the chances of success, where the $\sigma$ was used to
offset a $J$ value (from the objective function) that was too high
(overflow problems). Care should be taken to set both the initial
$\lambda_{k}$ and $\sigma$ to avoid problems.

\subsection{Vectorization and ECVA}

The features focused on, as part of our implementation, include both
vector variate and matrix variate views. The matrix variate views
requires transformation from their matrix domain to a vectorized domain
for implementation in the LM\textsuperscript{3}L framework. The matrix-variate
to vector-variate transformation implemented here is outlined in \citet{johnston2017variable}.
The matrix is transformed $\mathrm{vec}(X_{i}^{k})=x_{i}^{k}$ to
a vector domain. A dimensionality reduction process is implemented
as some of the matrices are large enough to result in large sparse
vectors (i.e., $20\times20$ DF matrix = 400 element vector). To reduce
the large sparse feature vector resulting from the unpacking of matrix,
we applied a supervised dimensionality reduction technique commonly
referred to as extended canonical variate analysis (ECVA) \citep{Norgaard2006}. 

The methodology for ECVA has roots in principle component analysis
(PCA). PCA is a procedure performed on large multidimensional datasets
with the intent of rotating what is a set of possibly correlated dimensions
into a set of linearly uncorrelated variables \citep{Sch06}. The
transformation results in a dataset, where the first principle component
(dimension) has the largest possible variance. PCA is an unsupervised
methodology---known labels for the data being processed is not
taken into consideration---thus a reduction in feature dimensionality
will occur. While PCA maximizes the variance, it might not maximize
the linear separability of the class space. 

In contrast to PCA, Canonical Variate Analysis (CVA) does take class labels into considerations. The variation
between groups is maximized resulting in a transformation that benefits
the goal of separating classes. Given a set of data $\mathbf{x}$
with: $g$ different classes, $n_{i}$ observations of each class;
following \citet{Johnson1992}, the within-group and between-group
covariance matrix is defined as equations \ref{eq:20} and \ref{eq:21}
respectfully.

\begin{equation}
\mathbf{S}_{within}=\frac{1}{n-g}\sum_{i=1}^{g}\sum_{j=1}^{n_{i}}(x_{ij}-\bar{x}_{ij})(x_{ij}-\bar{x}_{i})'\label{eq:20}
\end{equation}

\begin{equation}
\mathbf{S}_{between}=\frac{1}{g-1}\sum_{i=1}^{g}n_{i}(x_{i}-\bar{x})(x_{i}-\bar{x})'\label{eq:21}
\end{equation}

Where $n=\sum_{i=1}^{g}n_{i}$, $\bar{x}_{i}=\frac{1}{n_{i}}\sum_{j=1}^{n_{i}}x_{ij}$,
and $\bar{x}=\frac{1}{n}\sum_{j=1}^{n_{i}}n_{i}x_{i}$. CVA attempts
to maximize the equation \ref{eq:22}.

\begin{equation}
J(\mathbf{w})=\frac{\mathbf{w}'\mathbf{S}_{between}\mathbf{w}}{\mathbf{w}'\mathbf{S}_{within}\mathbf{w}}\label{eq:22}
\end{equation}

The equation is solvable so long as $\mathbf{\mathbf{S}_{\mathit{within}}}$
is non-singular, which need not be the case, especially when analyzing
multicollinear data. When the case arises that the dimensions of
the observed patterns are multicollinear, additional considerations
need to be made. \citet{Norgaard2006} outlines a methodology, Extended Canonical Variate Analysis (ECVA),
for handling these cases in CVA. Partial least squares analysis \citep[PLS2, ][]{Wold1939} is used to solve the above linear equation, resulting
in an estimate of $\mathbf{w}$, and given that, an estimate of the
canonical variates (the reduced dimension set). The application of
ECVA to our vectorized matrices results in a reduced feature space
of dimension $g-1$, this reduced dimensional feature space, per view,
is then used in the LM\textsuperscript{3}L classifier.

\section{Derivation of Large Margin Multi-View Metric Learning with Matrix Variates }
\label{sec:derLM3L}
This objective design is solved using a gradient descent solver
operation. To enforce the requirements of $\mathbf{U}_{k}\succ0$
and $\mathbf{V}_{k}\succ0$, the metrics are decomposed---$\mathbf{U}_{k}=\mathbf{\Gamma}_{k}^{T}\mathbf{\Gamma}_{k}$
and $\mathbf{V}_{k}=\mathbf{N}_{k}^{T}\mathbf{N}_{k}$. The
gradient of the objective function with respect to the decomposed
matrices $\mathbf{\Gamma}_{k}$ and $\mathbf{N}_{k}$ is estimated. The unconstrained optimum is found using the gradient
of the decomposed matrices; the $\mathbf{U}_{k}$ and $\mathbf{V}_{k}$ matrices are then reconstituted at the end of the optimization process. We reformulate the matrix variate distance as equation \ref{eq:34}:

\begin{equation}
d_{\mathbf{\Gamma}_{k},\mathbf{N}_{k}}(\Delta_{ij}^{k})=\mathrm{tr}\left[\mathbf{\Gamma}_{k}^{T}\mathbf{\Gamma}_{k}\left(\Delta_{ij}^{k}\right)^{T}\mathbf{N}_{k}^{T}\mathbf{N}_{k}\left(\Delta_{ij}^{k}\right)\right]\label{eq:34};
\end{equation}

for ease we make the following additional definitions: $d_{\mathbf{U}_{k},\mathbf{V}_{k}}(X_{i}^{k},X_{j}^{k})=d_{ij}^{k}$,
$X_{i}^{k}-X_{j}^{k}=\Delta_{ij}^{k}$, $\mathbf{A}_{ij}^{k}=\left(\Delta_{ij}^{k}\right)^{T}\mathbf{N}_{k}^{T}\mathbf{N}_{k}\left(\Delta_{ij}^{k}\right)$
, and $\mathbf{B}_{ij}^{k}=\Delta_{ij}^{k}\mathbf{\Gamma}_{k}^{T}\mathbf{\Gamma}_{k}\left(\Delta_{ij}^{k}\right)^{T}$.
Note that $\mathbf{A}_{ij}^{k}=\left(\mathbf{A}_{ij}^{k}\right)^{T}$
and $\mathbf{B}_{ij}^{k}=\left(\mathbf{B}_{ij}^{k}\right)^{T}$. Additionally
we identify the gradients as equations \ref{eq:35} and \ref{eq:36}:

\begin{equation}
2\mathbf{\Gamma}_{k}\mathbf{A}_{ij}^{k}=\frac{\partial d_{ij}^{k}}{\partial\mathbf{\Gamma}_{k}}\label{eq:35}
\end{equation}

\begin{equation}
2\mathrm{\mathbf{N}}_{k}\mathbf{B}_{ij}^{k}=\frac{\partial d_{ij}^{k}}{\partial\mathbf{N}_{k}}\label{eq:36},
\end{equation}

as being pertinent for derivation. We give the gradient
of the individual view objective $I_{k}$ as equations
\ref{eq:37} and \ref{eq:38}:

\begin{multline}
\frac{\partial I_{k}}{\partial\mathbf{\Gamma}_{k}}=2\mathbf{\Gamma}_{k} \\ \left(\left(1-\gamma\right)\sum_{i,j}\eta_{ij}^{k}\cdot\mathbf{A}_{ij}^{k}+ \gamma\sum_{j\rightsquigarrow i,l}\eta_{ij}^{k}\left(1-y_{il}\right)\cdot h'[z]\cdot\left[\mathbf{A}_{ij}^{k}-\mathbf{A}_{il}^{k}\right]+\lambda\mathbf{I} \right)\label{eq:37}
\end{multline}

\begin{multline}
\frac{\partial I_{k}}{\partial\mathbf{N_{k}}}=2\mathrm{\mathbf{N}}_{k} \\ \left(\left(1-\gamma\right)\sum_{i,j}\eta_{ij}^{k}\cdot\mathbf{B}_{ij}^{k}+\gamma\sum_{j\rightsquigarrow i,l}\eta_{ij}^{k}\left(1-y_{il}\right)\cdot h'[z]\cdot\left[\mathbf{B}_{ij}^{k}-\mathbf{B}_{il}^{k}\right]+\lambda\mathbf{I}\right)\label{eq:38},
\end{multline}

and the gradient of the joint objective as equations \ref{eq:39} and \ref{eq:40}:

\begin{equation}
\frac{\partial J_{k}}{\partial\mathbf{\Gamma}_{k}}=w_{k}^{p}\frac{\partial I_{k}}{\partial\mathbf{\Gamma}}_{k}+4\mu\mathbf{\Gamma}_{k}\sum_{q=1,q\neq k}^{K}\sum_{i.j}\left(d_{ij}^{k}-d_{ij}^{q}\right)\mathbf{A}_{ij}^{k}\label{eq:39}
\end{equation}

\begin{equation}
\frac{\partial J_{k}}{\partial\mathbf{N_{k}}}=w_{k}^{p}\frac{\partial I_{k}}{\partial\mathbf{N}}_{k}+4\mu\mathrm{\mathbf{N}_{k}}\sum_{q=1,q\neq k}^{K}\sum_{i.j}\left(d_{ij}^{k}-d_{ij}^{q}\right)\mathbf{B}_{ij}^{k}\label{eq:40}.
\end{equation}

To estimate the update for the weights, we solve for the Lagrange function given equation \ref{eq:41}:

\begin{multline}
La(w,\eta)=\sum_{k=1}^{K}w_{k}^{p}I_{k}+ \\
\lambda\sum_{k,l=1,k<l}^{K}\sum_{i,j}\left(d_{ij}^{k}-d_{ij}^{l}\right)^{2}-\eta\left(\sum_{k=1}^{K}w_{k}-1\right)\label{eq:41};
\end{multline}

we estimate the weights as equation \ref{eq:42}:

\begin{equation}
w_{k}=\frac{\left(1/I_{k}\right)^{1/(p-1)}}{\sum_{k=1}^{K}\left(1/I_{k}\right)^{1/(p-1)}}\label{eq:42}.
\end{equation}

The implementation of distance in the multi-view case, i.e. implementation
of distance used in the k-NN algorithm, is given as equation \ref{eq:Distance}:

\begin{equation}
d(X_{i},X_{j})=\sum_{k=1}^{K}w_{k}\mathrm{tr}\left[\mathbf{U}_{k}\left(X_{i}^{k}-X_{j}^{k}\right)^{T}\mathbf{V}_{k}\left(X_{i}^{k}-X_{j}^{k}\right)\right]\label{eq:Distance}
\end{equation}

We note the following about the algorithm: 
\begin{enumerate}
\item Similar to LM\textsuperscript{3}L we optimize in two stages at each iteration:
freezing the weights and optimizing $\mathbf{\Gamma}_{k}$ and $\mathbf{N}_{k}$
with respect to the primary objective function, then freezing the
estimates of $\mathbf{\Gamma}_{k}$ and $\mathbf{N}_{k}$ and optimizing
$w_{k}$ given the Lagrangian
\item The generation of the gradient for the objective is $\nabla J_{k}\left(X_{i}^{k};\mathbf{U}_{k},\mathbf{V}_{k}\right)=\left[\frac{\partial J_{k}}{\partial\Gamma_{k}},\frac{\partial J_{k}}{\partial\mathbf{N}_{k}}\right]$
; simultaneous estimate of the gradient is possible---there no need for flip-flopping the order of operation unlike the estimate of the sample covariance matrices themselves as shown in
\citet{glanz2013expectation}.
\item The step sizes for each iteration are estimated using our BB method
generated, step sizes for $\mathbf{U}_{k}$ and $\mathbf{V}_{k}$
are found independently from each other and from each view, i.e. the
equations \ref{eq:43} and \ref{eq:44}
\end{enumerate}

\begin{equation}
\hat{\beta}_{k}=\frac{1}{2}\cdot\frac{Tr\left[\Delta g(\mathbf{\Gamma}_{k})\cdot\Delta\mathbf{\Gamma}_{k}^{H}+\Delta\mathbf{\Gamma}_{k}\cdot\Delta g(\mathbf{\Gamma}_{k})^{H}\right]}{Tr\left[\Delta g(\mathbf{\Gamma}_{k})\cdot\Delta g(\mathbf{\Gamma}_{k})^{H}\right]}\label{eq:43}
\end{equation}

\begin{equation}
\hat{\kappa}_{k}=\frac{1}{2}\cdot\frac{Tr\left[\Delta g(\mathbf{N}_{k})\cdot\Delta\mathbf{N}_{k}^{H}+\Delta\mathbf{N}_{k}\cdot\Delta g(\mathbf{N}_{k})^{H}\right]}{Tr\left[\Delta g(\mathbf{N}_{k})\cdot\Delta g(\mathbf{N}_{k})^{H}\right]}\label{eq:44}
\end{equation}

The algorithm recombines the decomposed matrices to produce the results
$\mathbf{U}_{k}=\mathbf{\Gamma}_{k}^{T}\mathbf{\Gamma}_{k}$ and $\mathbf{V}_{k}=\mathbf{N}_{k}^{T}\mathbf{N}_{k}$
per view.

\section{Performance Comparison}
\label{sec:performanceComp}

Tables \ref{tab:LINEAR-Confusion-Matrix-Delta} and \ref{tab:UCR-Confusion-Matrix-Delta} contain delta values (differences) of LM\textsuperscript{3}L-MV - LM\textsuperscript{3}L confusion matrices for the datasets analyzed for this paper.

\begin{table*}
\caption{Difference LINEAR confusion matrix, LM\textsuperscript{3}L-MV - LM\textsuperscript{3}L\label{tab:LINEAR-Confusion-Matrix-Delta}}
\centering{}%
\begin{tabular}{|c|c|c|c|c|c||c|}
\hline 
Misclassification Rate & RR Lyr (ab) & Delta Scu / SX Phe & Algol & RR Lyr (c) & Contact Binary & Missed\tabularnewline
\hline 
\hline 
RR Lyr (ab) & -7 & 0 & 1  & 9 & -1 & -2\tabularnewline
\hline 
Delta Scu / SX Phe & 1 & 1 & 0 & 0 & -2 & 0\tabularnewline
\hline 
Algol & 2 & 0 & -4 & 1 & 1 & 0 \tabularnewline
\hline 
RR Lyr (c) & -1 & 0 & 0 & 2 & -1 & 0 \tabularnewline
\hline 
Contact Binary & 0 & 0 & 2 & -8 & 5 & 1 \tabularnewline
\hline 
\end{tabular}
\end{table*}

\begin{table*}
\centering{}
\caption{Difference UCR confusion matrix, LM\textsuperscript{3}L-MV - LM\textsuperscript{3}L \label{tab:UCR-Confusion-Matrix-Delta}}
\begin{tabular}{|c|c|c|c||c|}
\hline 
Misclassification Rate & 2 & 3 & 1 & Missed\tabularnewline
\hline 
\hline 
2 & 2 & -3 & 0 & 1 \tabularnewline
\hline 
3 & -13 & -171 & 184 & 0 \tabularnewline
\hline 
1 & -5 & 92 & -87 & 0 \tabularnewline
\hline 
\end{tabular} 
\end{table*}


\label{lastpage}
\end{document}